\documentclass[reprint,showpacs,preprintnumbers,amsmath,amssymb,prc,floatfix]{revtex4-1}
\usepackage{float}
\usepackage{color}
\usepackage{graphicx}
\usepackage{dcolumn}
\usepackage{bm}
\usepackage{rotating}
\usepackage{xcolor}
\usepackage{multirow}
\usepackage{adjustbox}
\usepackage{longtable}
\usepackage{comment}
\usepackage{natbib}
\usepackage{array}
\usepackage[normalem]{ulem}
\usepackage[colorlinks,citecolor=blue,linkcolor=red,anchorcolor=blue,filecolor=blue,urlcolor=blue]{hyperref}
\begin{document}
	
\title{Isospin dependent properties of the isotopic chain of Scandium and Titanium nuclei within the relativistic mean-field formalism}
	
\author{Praveen K. Yadav$^{1}$}
\email{praveenkumarneer@gmail.com}
\author{Raj Kumar$^{1}$}
\email{rajkumar@thapar.edu}
\author{M. Bhuyan$^{2,3}$}
\email{bunuphy@um.edu.my}
\affiliation{$^1$School of Physics and Materials Science, Thapar Institute of Engineering and Technology, Patiala-147004, Punjab, India}
\affiliation{$^2$Center for Theoretical and Computational Physics, Department of Physics, Faculty of Science, University of Malaya, Kuala Lumpur 50603, Malaysia}
\affiliation{$^3$Institute of Research and Development, Duy Tan University, Da Nang 550000, Vietnam}
	
\begin{abstract} 
\noindent
{\bf Background:} The density-dependent nuclear symmetry energy is directly related to the isospin asymmetry for finite and infinite nuclear systems. It is critical to determine the coefficients of symmetry energy and its related observables as it holds great importance in different areas of nuclear physics, such as analyzing the structure of ground state exotic nuclei and neutron star study. \\ 
{\bf Purpose:} The ground state bulk properties such as nuclear binding energy ($B.E.$), quadrupole deformation ($\beta_2$), two-neutron separation energy ($S_{2n} $), the differential variation of two-neutron separation energy ($dS_{2n} $), and root-mean-square charge radius ($r_{ch}$) for Scandium ($Z$ = 21), and Titanium ($Z$ = 22) nuclei are calculated. The isospin properties, namely the coefficient of nuclear symmetry energy and its components such as surface and volume symmetry energy for the finite isotopic chain from the corresponding quantities of infinite nuclear matter, are also estimated. Finally, we correlate the neutron-skin thickness with the coefficient of symmetry energy and its related observables corresponding to these isotopic chains of nuclei. \\ 
{\bf Methods:} The coherent density fluctuation model (CDFM) is used to estimate the isospin-dependent properties of finite nuclei such as symmetry energy, surface symmetry energy, and volume symmetry energy from its corresponding component in infinite nuclear matter. The relativistic mean-field (RMF) formalism with non-linear NL3 and Relativistic-Hartree-Bogoliubov with density-dependent DD-ME2 interaction parameters are employed in the present analysis. The weight function $\vert \mathcal{F}(x) \vert^{2} $ is estimated using the total density of each nucleus, which in turn is used along with the nuclear matter quantities to obtain the effective symmetry energy and its components in the finite nuclei.\\
{\bf Results:} We calculate the ground state bulk properties such as nuclear binding energy, quadrupole deformation, two-neutron separation energy, the differential variation of two-neutron separation energy, and root-mean-square charge radius for the Sc- and Ti- isotopic chains based on non-linear NL3 and density-dependent DD-ME2 parameter sets. Further, the ground state density distributions are used within the CDFM to obtain the effective surface properties such as symmetry energy and its components, namely volume and surface symmetry energy, for both the parameter sets. The calculated quantities are used to understand the isospin dependent structural properties of finite nuclei near and beyond the drip line that will further our horizon of finding newer magicity along the isotopic chains. \\
{\bf Conclusions:} A shape transition is observed from spherical to prolate near $N$ $\geq$ 44, and $N$ $\geq$ 40 for Sc- and Ti- isotopic chain, respectively. Notable signatures of shell and/or sub-shell closures have been found for the magic neutron numbers at $N$ = 20 and 28 for both the isotopic chain using the nuclear bulk and isospin quantities. In addition to these, a few signatures of shell/sub-shell closure are observed near drip-line region, at $N$ = 34 and 50 by following the surface/isospin dependent observables, namely symmetry energy and its component for both the isotopic chain of \textit{odd-A} Sc- and \textit{even-even} Ti- nuclei.
\end{abstract}

\maketitle
\noindent

\noindent
\section{Introduction}
Nuclei comprising large neutron-proton asymmetry that reside further away from the $ \beta $-stability line (exotic nuclei) are of great importance in understanding the modern nuclear structure. Mainly, there are three reasons which highlight the importance of exotic nuclei in the study of nuclear physics
: Firstly, to extend the data of nuclear structure as a function of $N$ and $Z$ in a well-ordered manner \cite{Otsuka2001}. Secondly, the nuclear structure phenomena such as the evolution of new shell closure and decay mode occurring far away from the $ \beta $-stability line are considerably different from those found on and near the $ \beta $-stability line \cite{Otsuka2010,Bhuyan2015}. Thirdly, it is crucial to the evolutionary study of the universe \cite{Meng2006}. The development of radioactive ion beam (RIB) facilities has created new avenues in the study of various nuclear phenomena, namely the island of inversion, proton radioactivity, bubble structure, Efimov effect, exotic shapes, and giant halo around the drip-line nuclei \cite{Mutschler2017,Singh2012,Ohnishi2008,Bhuyan2015,Jensen2004,Hansen2003,Pfuetzner2012,1955,Kaur2020}. Moving across the valley of stability towards the neutron drip-line, there appears to be an increase in the number of neutron-neutron pairs compared to neutron-proton pairs, resulting in a large amount of repulsive energy, which causes instability of the nuclei based on Pauli's principle. In this region, the shell effect overcomes the instability, leading to the shift of the drip-line further away and broadening of the peninsula of stability \cite{Satpathy2004}. 

In recent decades, substantial works was carried out to establish the shell evolution of neutron numbers $N$ = 32 and 34 along with the proton-magic chain of calcium ($Z$ = 20), rendering the newly emerged closed-shell phenomenon in this region. The basic mechanics corresponding to the appearance of $N$ = 32 and $N$ = 34 gaps for Ca is discussed in Ref. \cite{Otsuka2020}. It highlights the prediction of magicity based on the strong attractive coupling between the $\ell+1 / 2$ and $\ell-1 / 2$ orbits. In this $N$ = 32 and $N$ = 34 sub-shells, the protons ($\pi$) and neutrons ($\nu$) $p_{3/2}-p_{1/2}$ and $f_{7/2}-f_{5/2}$ spin-orbit interaction is used for the determination of the structure \cite{Xu2019}. A local maxima of the first $ 2^{+} $ excitation energies $ [E (2_{1}^{+})] $ were reported for the case of \textit{even-even} nuclei with $N$ = 32 isotopes of Ar ($Z$ = 18) \cite{PhysRevLett.114.252501}, Ca ($Z$ = 20) \cite{Huck_1985}, Sc ($Z$ = 21) \cite{LCTC2021}, Ti ($Z$ = 22) \cite{janssens2002structure,Michimasa2020,leistenschneider_dawning_2018}, and Cr ($Z$ = 24) \cite{prisciandaro2001new}, suggesting neutron shell closure at $N$ = 32. Also, a local minima of reduced transition probabilities B(E2;$ 0^{+}\rightarrow 2_{1}^{+} $) has inferred the existence of this sub-shell in Ti \cite{dinca2005reduced} and Cr \cite{burger2005relativistic} isotopes. A similar degree of doubly magic nature of $ ^{54}$Ca \cite{steppenbeck2013evidence} supply direct experimental evidence for the beginning of a substantial sub-shell closure isotope at $N$ = 34.

Furthermore, in  Ref. \cite{JLiuNiu2020}, it is observed that the relativistic Lagrangian with DD-ME2 \cite{Lalazissis2005}, PC-PK1 \cite{Zhao2010}, and PK series \cite{Long2004} and PKOi (i=1,2,3) \cite{Long2006,Long2008} parametrizations provide a dilute picture in predicting the magicity in $N$ = 32 and 34 Ca isotopes, whereas the relativistic Hartree-Fock (RHF) lagrangian PKA1 \cite{Long2007} that comprises the degree of freedom of $\rho$-tensor improves systematically the nuclear structural properties \cite{Long2009,Long2010,LJWang2013}. The strong coupling between the s$_{1/2}$ for both protons and neutrons with dissimilar principle numbers and neutron ($\nu$) $\nu$2p$_{1/2}$ orbits known as Dirac inversion partners play a significant role in the opening of $N$ = 32 and 34 subshells \cite{JJLi2016}. Parallelly various experimental advancement throughout the world, such as Jyaväskylä (Finland) \cite{Leino1995}, GSI and FAIR (Germany) \cite{Winkler2008,Geissel1992}, FRIB and ORNL (US) \cite{Thoennessen2010,Gross2000}, RIKEN (Japan) \cite{Sakurai2008}, GANIL (France) \cite{Mueller1991}, FLNR (Russia) \cite{Rodin2003}, and CSR (China) \cite{Sun2003} opens up new avenues in exploring the properties of exotic nuclei under the extreme conditions of isospin asymmetry. Here we have chosen the isotopic chain of Sc- and Ti- nuclei, as their atomic numbers are close to one of the doubly magic nuclei $^{40, 48}$Ca (that is $Z$ = 20 and $N$ = 20 and 28). Hence, the properties of the isotopic chains of these nuclei enable us to find the neutron-to-proton asymmetry and their effect on the properties of nuclei lying closer to the $\beta$-stability line and those of drip-line nuclei.

In principle, the ground-state properties are related to the nuclear bulk properties such as the binding energy, two neutron separation energy, single-particle spectra, shell correction energy, and so on, which predicts the traditional magic numbers and shell and/or sub-shell closure for the nuclei. However, moving away from the stability line near and/or beyond the drip-line, where isospin asymmetry is dominant over the traditional observables, makes it unfeasible to predict the shell/sub-shell closure using the energy domain \cite{Satpathy2004}. The dilute predictions of traditional observables allow us to consider the quantities dependent on the isospin asymmetry of the system. As seen in Ref. \cite{Otsuka2020}, we observe that the interplay of tensor- and central-force monopole contribute additively to the sharp increase of the $1 f_{5 / 2}$ effective single-particle energies (ESPE) relative to the $2 p_{3 / 2}$ and $2 p_{1 / 2}$ orbits.  This results in the appearance of (sub-shell closure) magic gap $N$ = 32 and $N$ = 34, following the shift of $1 f_{5 / 2}$ orbit above $2 p_{1 / 2}$. It is crucial to note here that the appearance of the (sub)magic gap $N$ = 32 is only present when $1 f_{5 / 2}$ lies between $2 p_{3 / 2}$ and $2 p_{1 / 2}$ and for $N$ = 34, this gap becomes larger as a consequence of tensor force. The appearance of $N$ = 34 vanishes when the effect of tensor-force is not considered owing to the linear-dependence of proton holes in $1 f_{7 / 2}$. To overcome this issue, one has to translate the isospin asymmetry dependent observable, so-called, symmetry energy of infinite nuclear matter to a finite nuclear system. In other words, owing to the abnormally high or low value of isospin asymmetry in the drip-line nuclei, the observables dependent on isospin asymmetry become crucial in understanding the broadening of the peninsula of stability and successfully predicting the magicity and shell and/or sub-shell closure in the finite nuclei.

The net asymmetry in stable nuclei is given as $I=(N-Z)/(N+Z)$ of range $\sim$ 0.24, which in terms of their baryon density has the form $\alpha = \left (\frac{\rho_n - \rho_p}{\rho_n + \rho_p} \right)$. Here $\rho_n$ and $\rho_p$ are the densities of neutron and proton respectively \cite{Steiner_2005}. The density type isospin asymmetry is described by isovector-vector $\rho$-meson exchange and the mass type asymmetry by isovector-scalar $\delta$-exchange \cite{Bhuyan2013,Kumar2018}. The nuclear symmetry energy is directly related to the isospin asymmetry for both finite and infinite nuclear systems. It plays an important role in varied area of nuclear physics, for example in the study of ground-state structural property of exotic nuclei \cite{Nik_i__2008,Van_Giai_2010,VAN_DALEN_2010}, the physics of giant collective excitation \cite{Rodin_2007,Biswal_2015}, reaction-dynamics of heavy-ion \cite{CHEN_2008,Colonna_2009,LI_2008}, dipole polarizability \cite{Reinhard_2010,Ivanovskaya_2011,Piekarewicz_2012,Roca_Maza_2013}, the mirror charge radii \cite{Brown_2017,Cai_2016} and neutron stars study \cite{Steiner_2005,Fattoyev_2012,Dutra_2012,Dutra_2014}. Nuclear symmetry energy can be defined as a measure of the energy gain in converting asymmetric nuclear matter to symmetric one. The symmetry energy depends on the density distribution of the protons and neutrons inside the matter. The increased interest in the study of the symmetry energy, coupled with its advantage in predicting the properties of finite and infinite nuclear matter, including nuclei near and beyond the drip-line, has led to the newer predictions and theoretical confirmation of the shell and/or sub-shell closure for isotopic and isotonic chains of nuclei lying across the nuclear landscape \cite{Bhuyan2018,Biswal2020,Kaur2020}. It is also observed that the neutron star radius is correlated to the density term associated with symmetry energy at saturation. The neutron skin size is determined by the relative strength of the symmetry energy between the peripheral region (less dense) and the central region (near saturation) \cite{Furnstahl2002,RocaMaza2011,Typel2001}. Thus, the neutron-skin thickness is defined as a density dependence of symmetry energy around the central (saturation) region. Theoretically, slope parameter ($ L $-coefficient) is correlated with neutron-skin thickness of $^{208}$Pb \cite{AlexBrown2000,Furnstahl2002,Centelles2009,RocaMaza2011} and neutron star radius \cite{Gaidarov2012}.

In the case of finite nuclei, the significant issue in constraining the symmetry energy is a well-defined surface with non-uniformity in the matter density distribution of the nucleons. The Liquid-Drop-Model is one of the more straightforward methods to evaluate the coefficients of symmetry energy for the finite nuclear system, with the coefficient of surface symmetry energy being directly proportional to $ A^{-1/3} $, where $ A $ is the mass number of the nucleus \cite{Agrawal2012}. The volume symmetry energy co-efficient is nearly independent of the shape of the nucleus \cite{Agrawal2012,Nikolov2011}. The coherent density fluctuation model (CDFM) is another effective approach, where surface properties of the nuclei have been analyzed by folding the nuclear matter properties within the Brueckner energy density functional in terms of the weight functions $\vert \mathcal{F}(x) \vert^2$ \cite{Antonov2004,Ivanov2008,Antonov2009,Bhuyan2018,Antonov2016,Quddus2020}. Recently for the isotopic chain of doubly close shell nuclei, Warda {\it et al.} have theoretically demonstrated that the stiffness of the symmetry energy is due to the bulk and surface constituents of neutron-skin thickness \cite{Warda2009,Warda2014}. Danielewicz had shown the correlation between the surface and volume symmetry energy and their ratio with the neutron-skin thickness \cite{danielewicz2003surface}. Furthermore, recently the temperature effect on the surface as well as bulk symmetry has been reported, which predicts that the surface symmetry energy is more sensitive to temperature than its volume part \cite{Lee2010,Agrawal2014}. 

In this present analysis, we estimate both the bulk nuclear properties such as binding energy, the binding energy per particle, charge radius and deformation, and the isospin-dependent properties such as symmetry energy, surface symmetry energy and volume symmetry energy for the \textit{odd-A} Scandium ($Z$ = 21) and \textit{even-even} Titanium ($Z$ = 22) isotopes within the relativistic mean-field (RMF) for non-linear NL3 \cite{Lalazissis1997}, and Relativistic-Hartree-Bogoliubov (RHB) with density-dependent meson-exchange DD-ME2 \cite{Lalazissis2005} parameter sets. The CDFM formalism is employed which allows the transition from the properties of the nuclear matter lying in the momentum space to that of the corresponding finite nuclear matter quantity lying in the coordinate space  \cite{Antonov2004,Ivanov2008,Antonov2009,Bhuyan2018,Antonov2016,Quddus2020}. This paper is organized in the following manner. In Section \ref{theory}, we present the theoretical formalism for the relativistic mean-field model followed by the coherent density fluctuation model (CDFM) \ref{CDFM}. Section \ref{Results} discusses the calculation and results obtained. Finally, a summary and conclusion are presented in Section \ref{Summary}. 

\section{Theoretical Formalism} \label{theory}
\noindent
In the present work, we evaluate the nuclear symmetry energy $S(\rho)$, surface symmetry energy $S_{S}$ and volume symmetry energy $S_{V} $ for exotic finite nuclei as a function of baryon density. The nuclear symmetry energy is an essential quantity in finite nuclei and infinite nuclear matter due to its isospin and density dependence behavior. The most general form of nuclear symmetry energy within relativistic mean-field formalism can be expressed as,
\begin{eqnarray}
	\centering 
	\label{eqn:1}
	S^{NM}(\rho)=\dfrac{1}{2}\dfrac{\partial^{2}(\mathcal{E}/\rho)}{\partial\alpha^{2}}\bigg|_{\alpha=0}.
\end{eqnarray}	
Here $\mathcal{E}$ corresponds to energy density and $\alpha$ is the neutron-proton asymmetry in terms of baryon density. Detailed expression for the symmetry energy co-efficient, namely the slope parameter $L^{NM}$, curvature $K_{0}^{NM}$ and skewness parameter $Q_{0}^{NM}$ can be found in Refs. \cite{Nik_i__2008,AlexBrown2000,boguta1977relativistic} and references therein. It is worth mentioning that this model has been widely applied to describe finite, infinite, and stellar nuclear matter in extreme isospin asymmetry \cite{Nik_i__2008,AlexBrown2000,boguta1977relativistic}.	

The relativistic mean-field (RMF) theory is one of the microscopic methods used to solve the many-body problem in nuclear physics. This theory is formulated within the framework of quantum hydrodynamics (QHD). RMF theory has several advantages compared to its non-relativistic counterpart; firstly, it naturally incorporates the spin-orbit force \cite{serot1992relativistic}. Secondly, it solves the shift of the saturation curve towards the empirical values (Coester band) \cite{brockmann1990relativistic}. Thirdly, it provides a successful description of finite nuclei on the stable and drip line regions and the super-heavy nuclei on the nuclear chart \cite{serot1997recent, fritz1993dirac, bodmer1991relativistic}. This model can predict properties of nuclei such as binding energy, root-mean-square (\textit{rms}) radius, nuclear density distributions, deformation parameter, single-particle energies of the ground, and their intrinsic excited states throughout the nuclear chart. In the present study, we have used the microscopic self-consistent RMF formalism to probe the nuclear structure. A characteristic relativistic Lagrangian density (generated after several changes to the original Walecka Lagrangian to account for its various limitations) for a nucleon–meson many-body systems has the form (Refs.\cite{Bhuyan2018,boguta1977relativistic,carlson2000dirac,lalazissis2009effective,nikvsic2002relativistic}):
\begin{eqnarray}
	{\cal L}&=&\overline{\psi}\{i\gamma^{\mu}\partial_{\mu}-M\}\psi +{\frac12}\partial^{\mu}\sigma
	\partial_{\mu}\sigma \nonumber \\
	&& -{\frac12}m_{\sigma}^{2}\sigma^{2}-{\frac13}g_{2}\sigma^{3} -{\frac14}g_{3}\sigma^{4}
	-g_{s}\overline{\psi}\psi\sigma \nonumber \\
	&& -{\frac14}\Omega^{\mu\nu}\Omega_{\mu\nu}+{\frac12}m_{w}^{2}\omega^{\mu}\omega_{\mu}
	-g_{w}\overline\psi\gamma^{\mu}\psi\omega_{\mu} \nonumber \\
	&&-{\frac14}\vec{B}^{\mu\nu}.\vec{B}_{\mu\nu}+\frac{1}{2}m_{\rho}^2
	\vec{\rho}^{\mu}.\vec{\rho}_{\mu} -g_{\rho}\overline{\psi}\gamma^{\mu}
	\vec{\tau}\psi\cdot\vec{\rho}^{\mu}\nonumber \\
	&&-{\frac14}F^{\mu\nu}F_{\mu\nu}-e\overline{\psi} \gamma^{\mu}
	\frac{\left(1-\tau_{3}\right)}{2}\psi A_{\mu},
	\label{lag}
\end{eqnarray}
with vector field tensors
\begin{eqnarray}
	F^{\mu\nu} = \partial_{\mu} A_{\nu} - \partial_{\nu} A_{\mu} \nonumber \\
	\Omega_{\mu\nu} = \partial_{\mu} \omega_{\nu} - \partial_{\nu} \omega_{\mu} \nonumber \\
	\vec{B}^{\mu\nu} = \partial_{\mu} \vec{\rho}_{\nu} - \partial_{\nu} \vec{\rho}_{\mu}.
\end{eqnarray}
Here the fields for the $\sigma$-, $\omega$- and isovector $\rho$- meson is denoted by $\sigma$, $\omega_{\mu}$, and $\vec{\rho}_{\mu}$, respectively. The electromagnetic field is defined by $A_{\mu}$. The quantities, $\Omega^{\mu\nu}$, $\vec{B}_{\mu\nu}$, and $F^{\mu\nu}$ are the field tensors for the $\omega^{\mu}$, $\vec{\rho}_{\mu}$ and photon fields, respectively.

The RMF formalism permits density dependence of the meson-nucleon coupling, which is parameterized in the phenomenological approach. More details can be found in the Refs.\cite{Bhuyan2018,boguta1977relativistic,carlson2000dirac,lalazissis2009effective,nikvsic2002relativistic}. From the above Lagrangian density, we obtain the field equations for the nucleons and the mesons by expanding the upper and lower components of the Dirac spinors and the boson fields in an axially deformed harmonic oscillator basis for an initial deformation $\beta_{0}$. The set of coupled equations is solved numerically by a self-consistent iteration method. The center-of-mass-energy correction is estimated using the formula $E_{c.m.} = \frac{3}{4}(41A^{-1/3})$ MeV. One can obtain the centre-of-mass-energy from macroscopic and microscopic self-consistent methods \cite{Lalazissis2005,Bhuyan2015}. The recent study of Bhuyan \cite{Bhuyan2015} shows that the centre-of-mass-energies of both methods almost overlap; hence the use of the macroscopic method is appropriate and also does not affect the final predictions. The total binding energy and other observable are also obtained using the standard relations, given in Ref. \cite{Gambhir1990}.

For studying open-shell nuclei, it is crucial to consider the pairing correlation in the ground and excited states. There are various methods, such as the BCS approach, the Bogoliubov transformation, and the particle number conserving methods, that have been developed to treat pairing correlation in the finite nuclei \cite{nikvsic2002relativistic, Lalazissis2005, Goriely2001, Lalazissis1997, Bhuyan2018, Gambhir1990, lalazissis2009effective}. In the present study, two different kinds of prescriptions are used for pairing correlations i.e., the BCS method with constant gap for NL3, while the Bogoliubov transformation with DD-ME2 \cite{ nikvsic2002relativistic,Lalazissis2005,Pannert1987,Goriely2001,Lalazissis1997,Bhuyan2018,Gambhir1990,lalazissis2009effective}. The reason for considering two types of prescription here is to investigate the effects of pairing as well as their model dependence on nuclear matter quantities at local density. For the nuclei lying close to the $\beta$-stability line, the constant gap BCS pairing approach can be used to obtain a reasonably good approximation of pairing \cite{Dobaczewski}. Whereas, it is found that in the case of nuclei lying in the exotic region, i.e., close to the drip-line are found to show a minimal effect when employing the BCS approach. This paves the way for the Bogoliubov transformation, which is found to be a viable method for dealing with pairing correlation in exotic drip-line nuclei \cite{Nik2002}. The Relativistic-Hartree- Bogoliubov (RHB) energy density functional comprises RMF functional along with the pairing of the RHB model and the pairing tensor \cite{Sahoo2019}. In the RHB prescription, energy is expressed as:
\begin{equation}	
\begin{aligned}
	& E_{\text {pair }}[\hat{k}]\\
	& =\frac{1}{4} \sum_{n_{1} n_{1} / n_{2} n_{2} \prime} \sum_{n_{1} n_{1}^{\prime}}^{*}<n_{1} n_{1} \prime\left|V^{P P}\right| n_{2} n_{2}^{\prime}>K_{n_{2} n_{2} \prime},
\end{aligned}
\end{equation}
where  $<n_{1} n_{1} \prime\left|V^{P P}\right| n_{2} n_{2}^{\prime}>$ refers to the two-body pairing matrix. The RHB equation taken the form as,
\begin{eqnarray}
	\left[\begin{array}{cc}
		\hat{h}_{D}-m-\lambda & \Delta \\ 
		-\Delta^{*} & -\hat{h}_{D}+m+\lambda
	\end{array}\right]\left[\begin{array}{c}
    	u_{k}(r) \\
		v_{k}(r)
	\end{array}\right] \nonumber \\
    	=E_{k}\left[\begin{array}{c}
		u_{k}(r) \\
		v_{k}(r)
	\end{array}\right].
\end{eqnarray}
Here $\hat{h}_{D}$ refers to single nucleon Dirac Hamiltonian, $m$ signify nucleonic mass, $\lambda$ refers to the chemical potential and $E_{k}$ signify quasi-particle energy eigenfunction. The pairing field can be represented as 
\begin{eqnarray}
	\left[\begin{array}{ll}
		\Delta_{++} & \Delta_{+-} \\
		\Delta_{-+} & \Delta_{--}
	\end{array}\right].
\end{eqnarray}
Even though the terms $\Delta_{-+}$ and $\Delta_{+-}$ contain many large terms, they provide minute contribution when compared with $\sigma \triangledown$ term of Dirac Hamiltonian \cite{Berger1984, Gonzales1996}. This led to the introduction of Gogny force in the pairing correlation \cite{Nik2006,Nik_i__2008}.

{\bf \subsection{Coherent density fluctuation model}}
\label{CDFM}
The coherent density fluctuation model (CDFM) was developed and formulated by Antonov {\it et al.} \cite{antonov1979model,antonov1982spectral}. It is a natural extension of Fermi-gas model and is based upon the $ \delta $-function limit of the generator coordinate method \cite{PhysRev.108.311}. It serves us to transform the various quantities from infinite nuclear matter to its corresponding finite one. Thus, using the CDFM one can define the symmetry energy for a finite nucleus by weighting the corresponding quantities of infinite nuclear matter at local denaity. Following the CDFM approach, the effective symmetry energy $S$ is given as \cite{Antonov1994,Brockmann1992, Gaidarov2011, Gaidarov2012, Fuchs1995, Sarriguren2007,Bhuyan2018}:
\begin{eqnarray}
	\label{eqn:18}
	S = \int_{0}^{\infty}dx \vert\mathcal{F}(x)\vert^{2}S^{NM}(x).
\end{eqnarray}
The term $\vert\mathcal{F}(x)\vert^2$ is a weight function which is given by the expression
\begin{eqnarray}
	\centering 
	\label{eqn:17}
	\vert \mathcal{F}(x) \vert^{2} = -\bigg(\frac{1}{\rho_{0}(x)}\frac{d\rho(r)}{dr} \bigg)_{r=x},
\end{eqnarray}
with the normalization as $\int_{0}^{\infty}dx \vert \mathcal{F}(x) \vert^{2} = 1$. Here, $\rho_{0}(x)=\dfrac{3A}{4\pi x^{3}} $ and $x$ is the spherical radius generator coordinate for all $ A $ nucleons inside the uniformly distributed spherical Fermi gas. The detailed analytical derivation for obtaining the density-dependent weight function can be followed in Refs. \cite{antonov1979model,antonov1982spectral,Antonov1994, Brockmann1992,Fuchs1995}.
The surface $S_{S}$ and volume $S_{V}$ components of symmetry energy are calculated separately using the Danielewicz's method over the liquid drop model \cite{danielewicz2003surface, Danielewicz2006}. The individual components, namely $S_{V}$ and $S_{S}$ can be calculated from the symmetry energy expression as,
\begin{eqnarray}
	\centering 
	\label{eqn:20}
	S_{V}=S\bigg(1+\dfrac{1}{\kappa A^{1/3}}\bigg)
\end{eqnarray}
and    
\begin{eqnarray}
	\centering 
	\label{eqn:21}
	S_{S}=\dfrac{S}{\kappa}\bigg(1+\dfrac{1}{\kappa A^{1/3}}\bigg),
\end{eqnarray}
respectively. Here, the term $\kappa$=$S_{V}/S_{S}$ which is calculated from the expression, 
\begin{eqnarray}
	\centering 
	\label{eqn:29}
	\kappa=\dfrac{3}{R\rho_{0}}\int_{0}^{\infty}dx \vert\mathcal{F}(x) \vert^{2} x \rho_{0}(x) \bigg[\bigg(\dfrac{\rho_{0}}{\rho(x)}\bigg)^{\gamma}-1\bigg],
\end{eqnarray}  
where $ \rho_{0} $ is the nuclear matter equilibrium density and $ R=r_{0}A^{1/3} $ \cite{Dieperink2007}. In the present calculations, we use $\gamma=0.3$ and the motive for selecting this is detailed in the results and discussion (Sec. \ref{Results}). \\ 

\section{Calculations and Results}
\label{Results}
In the relativistic mean-field (RMF) approach, the field equations are solved self-consistently by taking different inputs of the initial deformation $\beta_0$  \cite{Pannert1987,Goriely2001,Lalazissis1997,Bhuyan2018,Gambhir1990,lalazissis2009effective}. The desired number of major shells for fermions and bosons is $N_F$ = $N_B$ = 12, for the convergence of the ground state solutions in the considered mass region. The number of mesh points for Gauss-Hermite and Gauss-Laguerre integration are 20 and 24, respectively. We obtain the bulk properties from self-consistent RMF formalism, namely binding energy, binding energy per particle, charge radius, and deformations in the present study. We then utilize the mean-field matter densities and nuclear matter quantities at saturation to calculate the nuclear symmetry energy, volume and surface symmetry energy along with the ratio of volume and surface symmetry energy for the isotopic chains of \textit{odd-A} Scandium ($Z$ = 21) and \textit{even-even} Titanium ($Z$ = 22) nuclei. The coherent density fluctuation model (CDFM) is used to construct the bridge between the infinite nuclear matter and finite nuclei in terms of symmetry energy within the non-linear NL3  \cite{Lalazissis1997} and density-dependent meson-exchange (DD-ME2) \cite{Lalazissis2005} effective interaction parameter sets. It is worth mentioning that these parameters are widely successful in providing a fairly good account of the properties from light to super-heavy nuclei across the neutron-proton drip line \cite{Bhuyan2015,Patra2009,Quddus2020,Naz2019}. \\

\begin{figure}[htpb]
	\includegraphics[scale=0.46]{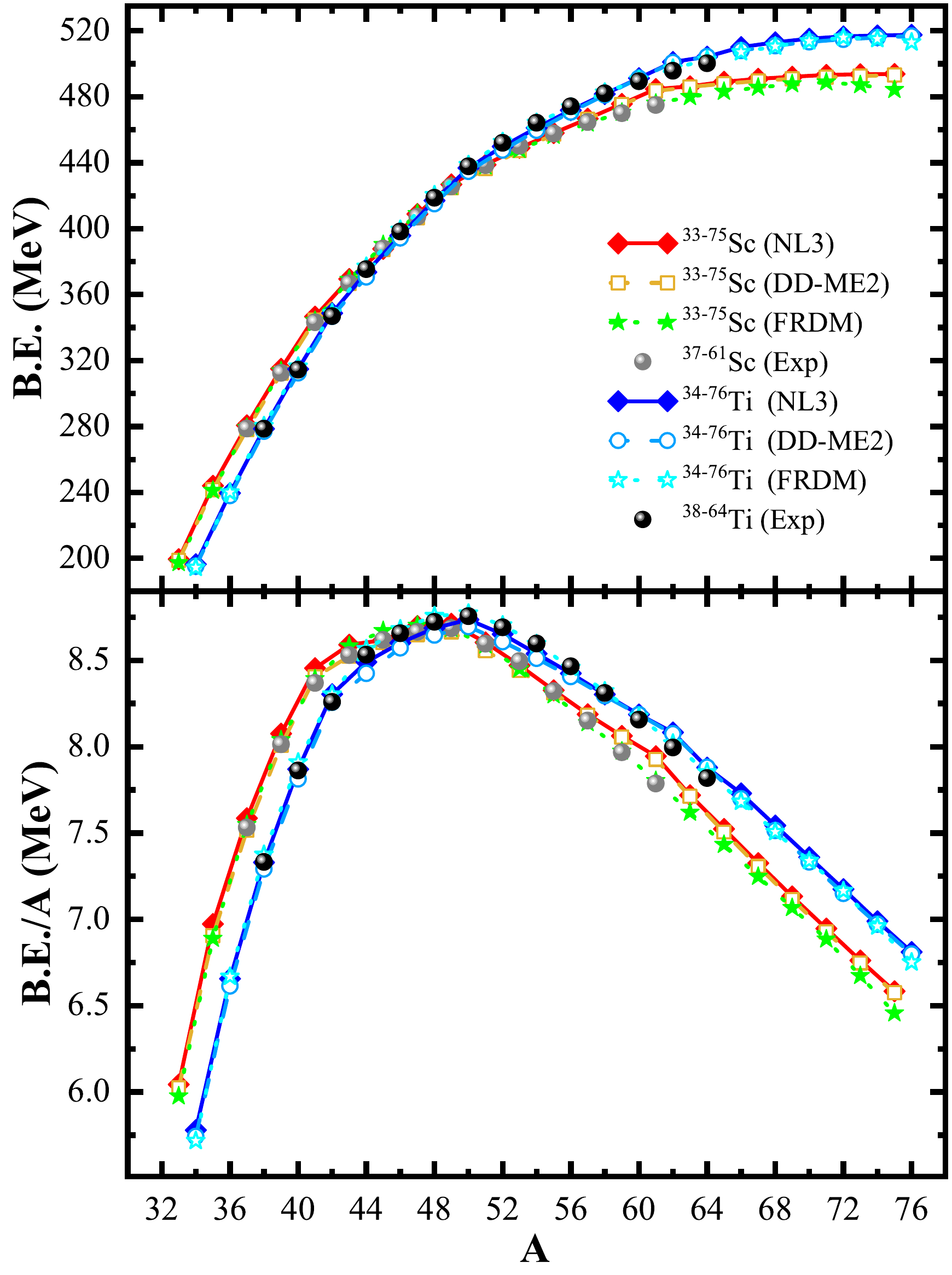}
	\caption{The binding energy ($B.E.$) and the binding energy per nucleon ($B.E./A$) from the non-linear NL3 and density-dependent DD-ME2 interaction for Sc and Ti isotopic chains as a function of mass number ($A$) are shown in the upper and lower panel, respectively. The FRDM predictions \cite{Moeller2016} and available experimental data \cite{Wang2012} are given for comparison.}
	\label{fig1}
\end{figure}

{\bf \subsection{Nuclear bulk properties}}
\noindent
{\bf Binding Energy and Pairing Energy:} Nuclear binding energy ($B.E.$) is the most fundamental and precise measured observable, facilitating a profound perspective in determining the shell/sub-shell of nuclei over the isotopic and isotonic chain. The efficiency of a theoretical model is determined from its consistency in generating accurate experimental binding energy. The binding energy for \textit{odd-A} Scandium ($Z$ = 21) and \textit{even-even} Titanium ($Z$ = 22) isotopic chain are calculated from relativistic mean-field for non-linear NL3 and Relativistic-Hartree-Bogoliubov for density-dependent DD-ME2 interaction parameter sets. The results are listed in Table \ref{tab1} and \ref{tab2}; and also shown in the upper panel of Fig. \ref{fig1} along with the FRDM predictions \cite{Moeller2016}, and the experimental data \cite{Wang2012}. The standard deviation of the RMF results obtained for NL3 and DD-ME2 parameter sets and FRDM predication with respect to the available data in the revised final version of the manuscript. The magnitude of the standard deviation of NL3, DD-ME2 parameter sets, and FRDM predictions is 3.66, 3.16, and 1.43 MeV for the Sc- isotopes and  2.47, 3.41, and 1.34 MeV for the Ti- isotopes, respectively. One can notice that the NL3 and DD-ME2 have higher standard deviations with respect to FRDM prediction, which is owed to the self-consistent microscopic models \cite{Gambhir1990,Lalazissis1999,lalazissis2009effective}.

The binding energy per particle ($B.E./A$) for both the isotopic chains is shown in the lower panel of Fig. \ref{fig1}. From the figure and Table \ref{tab1} and \ref{tab2}, it can be observed that the results of our calculations for binding energy are of reasonably good accuracy as it is overlapping with the experimental data and FRDM predictions, including the neutron-rich region of the isotopic chains. A detailed inspection shows that the value of binding energy per particle starts increasing with the neutron number, reaching a peak value at $N$ = 28 for both the chain and then finally decreases. Further careful observation in the neutron-rich side of the isotopic chain shows minute discontinuity or kink at $N$ = 40, which is one of the primary interests of the present analysis. Interestingly, no evidence of discontinuity is observed at the traditional magic number $N$ = 50. It is worth mentioning that the kink in the binding energy per particle exhibits greater stability of the isotope, i.e., a possible shell/sub-shell closure of the nuclei, compared to its neighbouring isotopes. From Table \ref{tab1} and \ref{tab2} one can observe that in the case of the NL3 parameter set, the pairing energy (E$_{pair}$) decreases over both Sc- and Ti- isotopic chains. In contrast, the DD-ME2 parameter shows relatively large abnormal changes in the pairing energy over both the isotopic chains. A careful examination of the relative variation of the magnitude of E$_{pair}$ shows sharp discontinuities/kinks at specific neutron numbers, namely $N$= 20, 28 and 40 for Sc- nuclei and at $N$= 20 and 40 for Ti nuclei in the case of the DD-ME2 parameter set, which is in stark contrast with the minute magnitude obtained in the case of the NL3 parameter set. This method provides a better insight into the trend in pairing energy from the Bogoliubov transformation over the BCS approach \cite{Nik2002,Nik2006}.
\begin{figure}[htpb]
	\includegraphics[width=8.7cm,height=7.5cm]{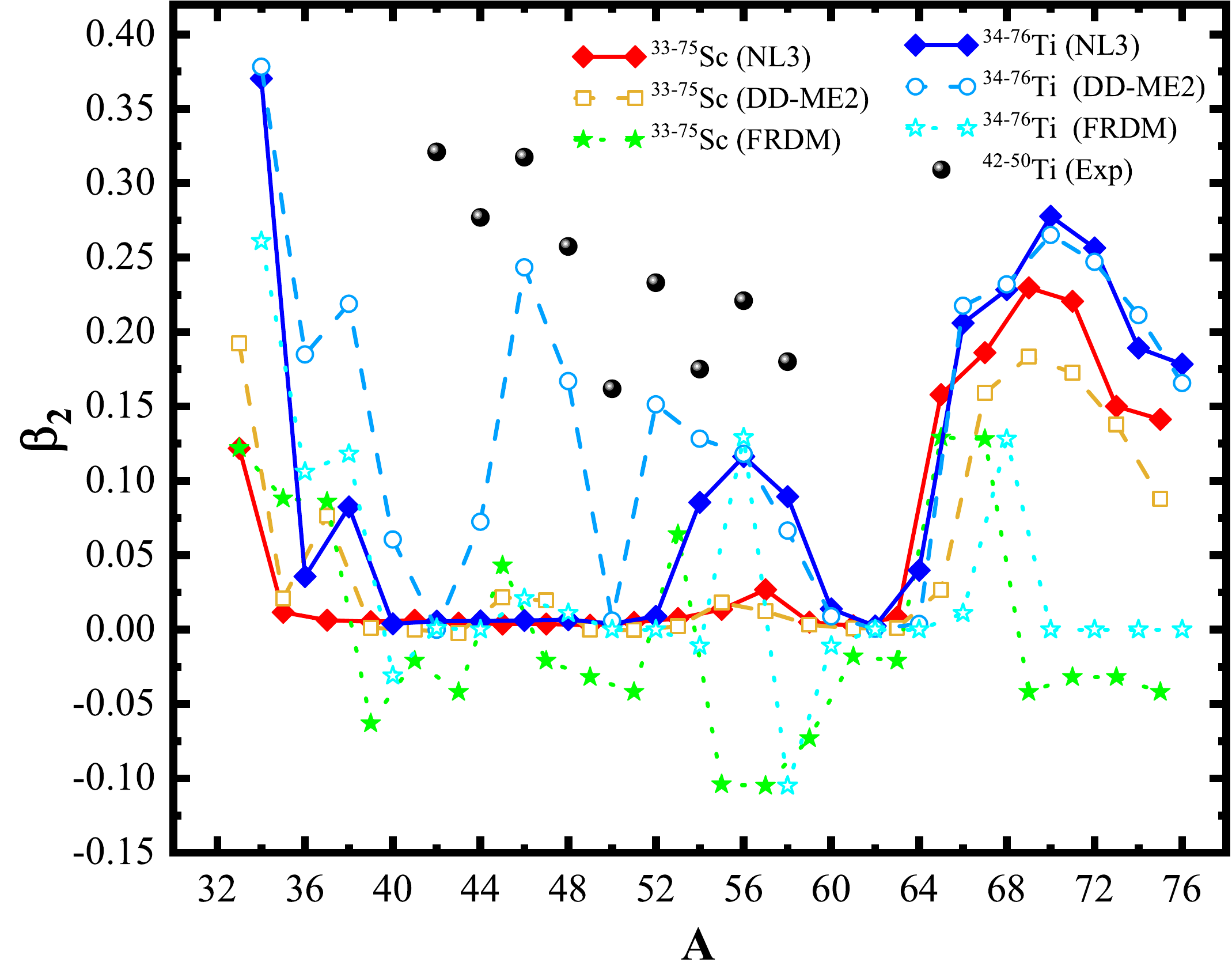}
	\caption{The quadrupole deformation $\beta_{2}$ from the NL3 and DD-ME2 interaction parameter sets for Sc and Ti isotopic chains as a function of mass number ($A$) are compared with the FRDM predictions \cite{Moeller2016} and available experimental data \cite{Pritychenko2016}.}
	\label{fig2}
\end{figure}
\begin{figure}[htpb]
	\includegraphics[scale=0.46]{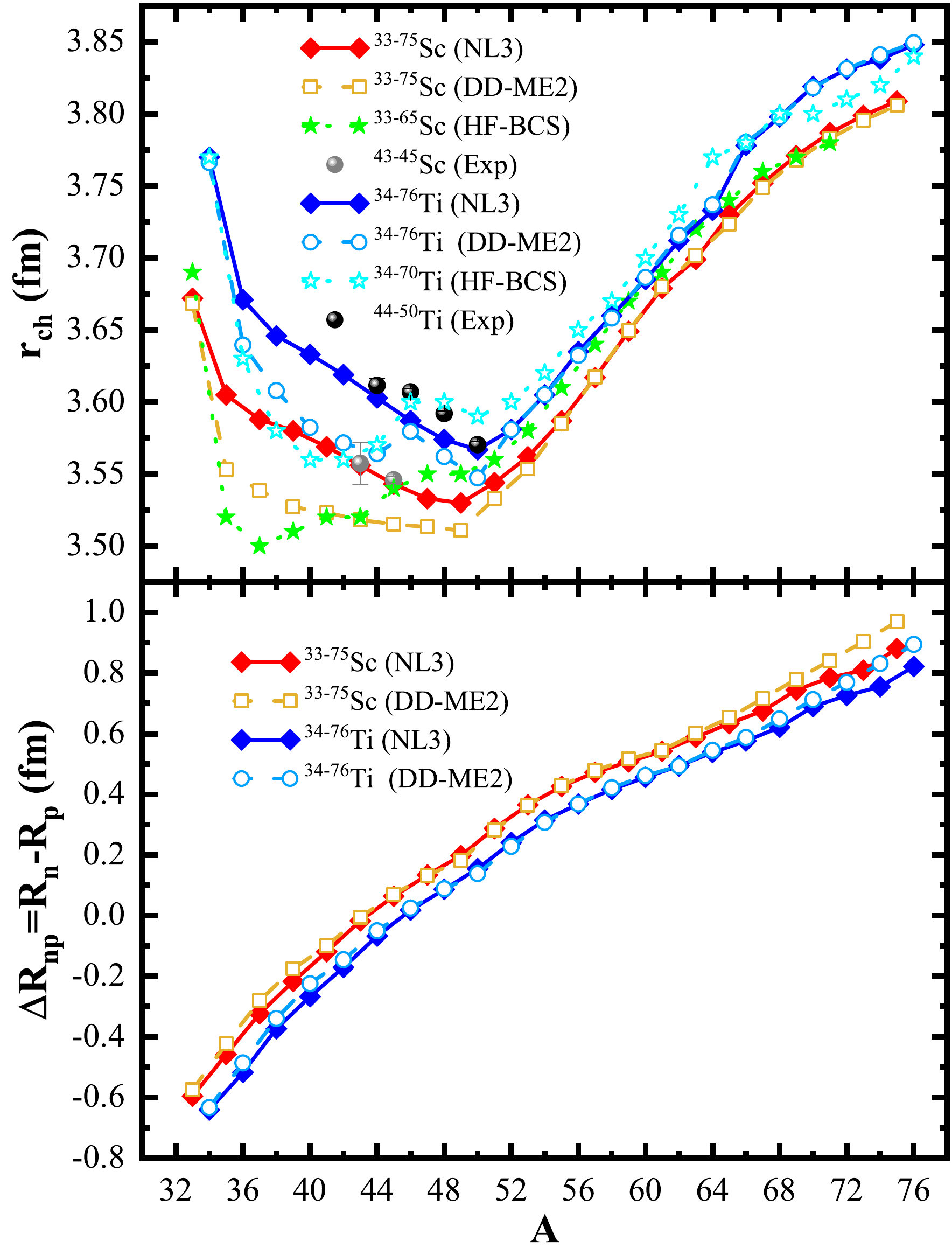}
	\caption{The charge radius $r_{ch}$ and neutron-skin thickness $ \Delta R_{np} $ from the NL3 and DD-ME2 interaction as a function of mass number ($A$) for Sc and Ti isotopic chains. The Hartree–Fock+Bardeen–Cooper–Schrieffer (HF-BCS) predictions  \cite{Goriely2001} and available experimental data \cite{Angeli2013} are given for comparison.}
	\label{fig4}
\end{figure}

\noindent
{\bf Quadrupole deformation:} The quadrupole deformation parameter $\beta_2$ is evaluated from the resulting proton and neutron quadrupole moments \cite{Gambhir1990}, as 
\begin{eqnarray}
Q = Q_n + Q_p =\sqrt{\frac{16\pi}5} \left (\frac3{4\pi} A R^2 \beta_2 \right).
\end{eqnarray}
The ground state $\beta_{2}$ is calculated from the RMF for NL3 and Relativistic-Hartree-Bogoliubov DD-ME2 parameter sets for the isotopic chain of Sc- and Ti- nuclei. In Fig. \ref{fig2}, we have shown the $\beta_{2} $ as a function of neutron number for \textit{odd-A} Scandium, and \textit{even-even} Titanium isotopic chains, using both the parameter sets along with the FRDM predictions \cite{Moeller2016} and experimental data \cite{Pritychenko2016}. In Fig. \ref{fig2}, we notice that the quadrupole deformation for the NL3 parameter set and FRDM predictions underestimate the experimental data, wherever available. In contrast, the DD-ME2 parameter estimate is quite close to the experimental data except for a few isotopes of Ti-nuclei. In FRDM, many nuclei are oblate in their ground state, whereas only a few oblate-shaped nuclei are observed for DD-ME2, while no such shape is observed for NL3 parameter set in both Scandium and Titanium isotopic chains. Moreover, we found a shape transition from spherical to prolate near $N$ $\geq$ 44 for Scandium and near $N$ $\geq$ 40 for Titanium isotopic chain in both the NL3 and DD-ME2 parameter sets. It is important to note here that the theoretical ground state quadrupole deformation is primarily characterized by shape fluctuations that construe quadrupole moment effectiveness as insignificant \cite{Poves2020}. Moreover, in many of the Sc- and Ti- nuclei, being not (axially) well-deformed, the experimental quadrupole deformation estimated from the experimental B (E2, 2+ $\rightarrow$ 0+) values assuming a rotational formula is ineffective. These calculations are provided only for the readers to get a firsthand comparison-based idea regarding the shape degrees of freedom of these nuclei. \\

\noindent
{\bf Charge radius and neutron-skin thickness:} Apart from the nuclear deformations, the root-mean-square (\textit{rms}) charge radius is also a significant quantity associated with the shape of finite nuclei. Further, the \textit{rms} radii of protons and neutrons can be used to impose the constraints not only on the saturation properties of infinite nuclear matter but also for studying the stellar collapse and type-II supernova explosions \cite{Buchinger1994}. Moreover, the charge radius provides structural effects corresponding to nuclear shape transitions and nuclear spectroscopy, which provide effective information about the shell of the nucleus \cite{Angeli2015}. The expression for root-mean-square (\textit{rms}) radius can be given as \cite{Patra1991, Bhuyan2015,BhuyanMahapatro,Gambhir1990}, 
\begin{equation}
	\centering 
	\label{eqn:33}
	<r^{2}_{m}>=\frac{1}{A}\int \rho(r_{\perp},z)r^{2}d\tau.
\end{equation}  
Here $A$ is the mass number, $\rho(r_{\perp}, z)$ is the axially deformed density and $\tau$ is the volume. From \textit{rms} proton radius, it is relatively easy to calculate the \textit{rms} charge radius by using the simple relation, $<r^{2}_{ch}>=<r^{2}_{p}>+0.64$ \cite{Wang2001}.
The macroscopic-microscopic models and effective interactions within microscopic mean-field formulations are widely in the calculation of the \textit{rms} charge radius. \cite{Wang2001,Angeli2013,Stoitsov2003,Goriely2010,Lalazissis1999,Zhao2010}. We have estimated the variation of charge radius over the isotopic chains of \textit{odd-A} Scandium and \textit{even-even} Titanium nuclei, as shown in the top panel of Fig. \ref{fig4}. The calculated charge radius ($r_{ch}$) for NL3 and DD-ME2 parameter sets are compared with the HF-BCS predictions \cite{Goriely2001}, and the latest experimental data \cite{Angeli2013}. From the figure, a uniform variation can be observed for both NL3 and DD-ME2 parameter sets. For example, a sharp fall in the charge radius from the proton-rich side of the isotopic chain for both Sc- and Ti- nuclei followed a depth at the neutron magic number $N$ = 28, and then uniformly increased with the mass number. A detailed inspection shows poor abnormality around $N$ = 40 for both the isotopic chains. Overall, the calculated values for NL3 and DD-ME2 are in good agreement with the HF-BCS \cite{Goriely2001} and experimental data \cite{Angeli2013}.

The neutron-skin thickness of a nucleus is defined as the difference between the \textit{rms} radius of the neutrons and protons inside the nucleus and expressed as $\Delta R_{np}=<r^{2}_{n}>-<r^{2}_{p}>$. Experimentally, the value of the proton \textit{rms} radius is obtained from the charge radius. The neutron skin-thickness is one of the direct key parameters from the finite nuclei connected with the isospin-dependent quantities in infinite nuclear matter, discussed in Sec. \ref{isospin}. For example, the mean-field models predict a linear correlation between $\Delta R_{np}$ of $^{208}$Pb and the slope of the symmetry energy of infinite nuclear matter at saturation density \cite{AlexBrown2000,Typel2001,Furnstahl2002,danielewicz2003surface,Warda2009,Avancini2007,Antonov2004,Ivanov2008,Antonov2009,Bhuyan2018,Antonov2016,Quddus2020}. Furthermore, in the finite nuclei, the $\Delta R_{np} $ is connected with the surface and volume symmetry energy, which can serve as one of the observable for determining the shell/sub-shell closure in nuclei near and beyond the drip-line \cite{danielewicz2003surface,Centelles2009, Satula2006,Antonov2004,Ivanov2008,Antonov2009,Bhuyan2018,Antonov2016,Quddus2020}. The variation of neutron-skin thickness over the isotopic chain of \textit{odd-A} Scandium and \textit{even-even} Titanium isotopes as a function of mass number is shown in the bottom panel of Fig. \ref{fig4}. From the figure, it can be observed that both NL3 and DD-ME2 parameter set show smooth increments with increasing mass numbers. Minor peaks are observed at neutron magic numbers $N$ = 20 and 28, whereas no appreciable change is observed at neutron magic number $N$ = 50. Moreover, little falls are observed at $N$ = 34 and 40, predicting the existence of shell/sub-shell closure. \\

\begin{table*}
\caption{The ground state binding energy, root-mean-square charge radius, quadrupole deformation and pairing energy for Sc isotopes of mass 33 {$\leqslant$} A {$\leqslant$} 75 are shown for RMF (NL3 and DD-ME2) parameter sets.  The available experimental data \cite{Wang2012,Angeli2013}, Finite-Range-Droplet-Model (FRDM) \cite{Moeller2016} and Hartree–Fock plus Bardeen–Cooper–Schrieffer (HF-BCS) \cite{Goriely2001} predictions are given for comparison. The energies are in MeV, and radii in fm.}
	\begin{adjustbox}{width=\textwidth}
		\renewcommand{\tabcolsep}{0.10cm}
		\renewcommand{\arraystretch}{1.27}
		\begin{tabular}{|c|cccc|cccc|ccc|cc|}
			\hline \hline 
			 Nucleus & \multicolumn{4}{c}{Binding Energy} & \multicolumn{4}{|c|}{Charge radius} & \multicolumn{3}{|c|}{Quadrupole deformation} & \multicolumn{2}{|c|}{Pairing Energy} \\
			 & NL3   & DD-ME2 & FRDM  & Expt. & NL3   & DD-ME2 & HF-BCS & Expt. & NL3   & DD-ME2 & FRDM  & NL3   & DD-ME2 \\
			 &       &       & \cite{Moeller2016} & \cite{Wang2012} &       &       & \cite{Goriely2001} & \cite{Angeli2013} &       &       & \cite{Moeller2016} &       &  \\
			\hline 
			$^{33} $Sc & 199.444 & 198.811 & 197.22 & -- & 3.672 & 3.6685 & 3.69  & -- & 0.1217 & 0.19224 & 0.122 & -8.193 & -5.794 \\
			$^{35} $Sc & 244.043 & 241.688 & 241.12 & -- & 3.605 & 3.553 & 3.52  & -- & 0.0115 & 0.02075 & 0.088 & -11.576 & -2.702 \\
			$^{37} $Sc & 280.696 & 278.104 & 279.25 & 278.684 & 3.588 & 3.5385 & 3.5   & -- & 0.0063 & 0.0765 & 0.086 & -15.973 & -6.146 \\
			$^{39} $Sc & 314.941 & 312.269 & 313.54 & 312.527 & 3.58  & 3.5272 & 3.51  & -- & 0.0053 & 0.0012 & -0.063 & -17.853 & -7.256 \\
			$^{41} $Sc & 346.66 & 344.522 & 344.09 & 343.137 & 3.569 & 3.5231 & 3.52  & -- & 0.0063 & -0.00023 & -0.021 & -17.786 & -3.089 \\
			$^{43} $Sc & 369.427 & 366.531 & 369.08 & 366.825 & 3.556 & 3.5181 & 3.52  & 3.5575 & 0.0043 & -0.00241 & -0.042 & -18.895 & -8.493 \\
			$^{45} $Sc & 387.786 & 387.128 & 390.21 & 387.852 & 3.543 & 3.5152 & 3.54  & 3.5459 & 0.0037 & 0.02165 & 0.043 & -17.893 & -9.842 \\
			$^{47} $Sc & 408.855 & 406.519 & 408.97 & 407.259 & 3.533 & 3.5133 & 3.55  & -- & 0.0038 & 0.01953 & -0.021 & -15.799 & -8.399 \\
			$^{49} $Sc & 426.902 & 424.604 & 425.49 & 425.627 & 3.53  & 3.5108 & 3.55  & -- & 0.003 & -0.00019 & -0.032 & -13.301 & -3.45 \\
			$^{51} $Sc & 438.783 & 436.365 & 437.98 & 438.437 & 3.544 & 3.533 & 3.56  & -- & 0.0049 & -0.0004 & -0.042 & -12.712 & -6.382 \\
			$^{53} $Sc & 448.925 & 447.332 & 447.86 & 450.256 & 3.562 & 3.5536 & 3.58  & -- & 0.0075 & 0.00214 & 0.064 & -11.193 & -6.835 \\
			$^{55} $Sc & 458.004 & 457.132 & 456.45 & 457.655 & 3.587 & 3.5852 & 3.61  & -- & 0.01355 & 0.01814 & -0.104 & -9.708 & -9.404 \\
			$^{57} $Sc & 466.716 & 466.378 & 463.88 & 464.607 & 3.617 & 3.6172 & 3.64  & -- & 0.0268 & 0.01222 & -0.105 & -8.175 & -10.068 \\
			$^{59} $Sc & 475.708 & 475.217 & 470.43 & 470.053 & 3.649 & 3.6493 & 3.67  & -- & 0.00505 & 0.00319 & -0.073 & -6.633 & -8.458 \\
			$^{61} $Sc & 484.672 & 483.537 & 475.94 & 475.007 & 3.679 & 3.6801 & 3.69  & -- & 0.00256 & 0.00074 & -0.018 & -4.939 & -3.707 \\
			$^{63} $Sc & 486.301 & 485.943 & 480.02 & -- & 3.699 & 3.7017 & 3.72  & -- & 0.00825 & 0.00101 & -0.021 & -5.16 & -9.554 \\
			$^{65} $Sc & 489.061 & 487.862 & 483.12 & -- & 3.73  & 3.7235 & 3.74  & -- & 0.15788 & 0.02663 & 0.129 & -3.548 & -12.57 \\
			$^{67} $Sc & 490.841 & 489.48 & 485.61 & -- & 3.752 & 3.749 & 3.76  & -- & 0.1862 & 0.15906 & 0.128 & -3.071 & -10.298 \\
			$^{69} $Sc & 492.145 & 490.869 & 487.64 & -- & 3.771 & 3.768 & 3.77  & -- & 0.22967 & 0.18334 & -0.042 & -2.58 & -11.152 \\
			$^{71} $Sc & 493.187 & 491.864 & 488.81 & -- & 3.787 & 3.7831 & 3.78  & -- & 0.22064 & 0.17257 & -0.032 & -2.119 & -12.539 \\
			$^{73} $Sc & 493.592 & 492.559 & 487.17 & -- & 3.799 & 3.7955 & -- & -- & 0.14992 & 0.13791 & -0.032 & -1.852 & -14.115 \\
			$^{75} $Sc & 493.746 & 493.1 & 484.31 & -- & 3.809 & 3.8059 & -- & -- & 0.14131 & 0.08787 & -0.042 & -1.641 & -15.204 \\
			\hline \hline
		\end{tabular}%
	\end{adjustbox}
	\label{tab1}%
\end{table*}%

\begin{table*}
	\caption{The ground state binding energy, root-mean-square charge radius, quadrupole deformation and pairing energy for Ti isotopes of mass 34 {$\leqslant$} A {$\leqslant$} 76 are shown for RMF (NL3 and DD-ME2) parameter sets.  The available experimental data \cite{Wang2012,Angeli2013,Pritychenko2016}, Finite-Range-Droplet-Model (FRDM) \cite{Moeller2016} and Hartree–Fock plus Bardeen–Cooper–Schrieffer (HF-BCS) \cite{Goriely2001} predictions are given for comparison. The energies are in MeV, and radii in fm.}
	\begin{adjustbox}{width=\textwidth}
		\renewcommand{\tabcolsep}{0.10cm}
		\renewcommand{\arraystretch}{1.27}
		\begin{tabular}{|c|cccc|cccc|cccc|cc|}
			\hline \hline 
			Nucleus & \multicolumn{4}{|c|}{Binding energy}  
			& \multicolumn{4}{|c|}{Charge radius} 
			& \multicolumn{4}{|c|}{Quadrupole deformation}
			& \multicolumn{2}{|c|}{Pairing Energy} \\
			& NL3 & DD-ME2 & FRDM & Expt. & NL3   & DD-ME2 & HF-BCS & Expt. & NL3 & DD-ME2 & FRDM & Expt. & NL3 & DD-ME2 \\
			& & & \cite{Moeller2016}  & \cite{Wang2012} & & & \cite{Goriely2001} & \cite{Angeli2013} & & & \cite{Moeller2016}  
			& \cite{Pritychenko2016} & & \\
			\hline 
			$^{34} $Ti & 196.5 & 195.259 & 194.38 & -- & 3.77  & 3.7662 & 3.77  & -- & 0.37039 & 0.37824 & 0.261 & -- & -6.129 & -2.018 \\
			$^{36} $Ti & 239.629 & 238.114 & 240.07 & -- & 3.671 & 3.6395 & 3.63  & -- & 0.03558 & 0.18489 & 0.106 & -- & -11.128 & -2.649 \\
			$^{38} $Ti & 278.536 & 277.068 & 280.22 & 278.616 & 3.646 & 3.6079 & 3.58  & -- & 0.08241 & 0.21885 & 0.118 & -- & -15.363 & -2.753 \\
			$^{40} $Ti & 314.811 & 312.505 & 316.36 & 314.48 & 3.633 & 3.5823 & 3.56  & -- & 0.00393 & 0.06044 & -0.031 & -- & -18.071 & -8.615 \\
			$^{42} $Ti & 348.785 & 346.784 & 348.74 & 346.888 & 3.619 & 3.5716 & 3.56  & -- & 0.00569 & -0.0002 & 0.001 & 0.319 & -18.384 & -4.967 \\
			$^{44} $Ti & 373.608 & 370.707 & 376.91 & 375.475 & 3.603 & 3.5642 & 3.57  & 3.6115 & 0.00588 & 0.07239 & 0     & 0.268 & -19.892 & -9.141 \\
			$^{46} $Ti & 395.692 & 394.309 & 399.29 & 398.197 & 3.587 & 3.5795 & 3.6   & 3.607 & 0.00608 & 0.24338 & 0.021 & 0.317 & -19.271 & -2.644 \\
			$^{48} $Ti & 417.018 & 415.046 & 420.36 & 418.704 & 3.574 & 3.5621 & 3.6   & 3.5921 & 0.00654 & 0.16686 & 0.011 & 0.269 & -17.071 & -5.662 \\
			$^{50} $Ti & 436.809 & 434.735 & 438.59 & 437.786 & 3.567 & 3.5473 & 3.59  & 3.5704 & 0.00399 & 0.00604 & 0     & 0.166 & -14.717 & -5.394 \\
			$^{52} $Ti & 449.875 & 447.628 & 452.6 & 451.967 & 3.581 & 3.5808 & 3.6   & -- & 0.0088 & 0.15133 & 0     & -- & -14.488 & -4.914 \\
			$^{54} $Ti & 461.131 & 459.622 & 464.16 & 464.26 & 3.605 & 3.6051 & 3.62  & -- & 0.08554 & 0.12822 & -0.011 & -- & -12.906 & -8.326 \\
			$^{56} $Ti & 471.776 & 470.734 & 473.65 & 474.102 & 3.635 & 3.6324 & 3.65  & -- & 0.11633 & 0.11806 & 0.129 & -- & -11.024 & -9.398 \\
			$^{58} $Ti & 481.569 & 481.084 & 482.65 & 482.038 & 3.66  & 3.6582 & 3.67  & -- & 0.08938 & 0.06648 & -0.105 & -- & -9.487 & -11.333 \\
			$^{60} $Ti & 491.237 & 491.159 & 490.44 & 489.42 & 3.685 & 3.6865 & 3.7   & -- & 0.01385 & 0.00885 & -0.011 & -- & -8.103 & -10.762 \\
			$^{62} $Ti & 501.136 & 500.647 & 497.16 & 495.69 & 3.712 & 3.7157 & 3.73  & -- & 0.00239 & 0.00056 & 0     & -- & -6.217 & -5.895 \\
			$^{64} $Ti & 504.292 & 504.322 & 502.54 & 500.352 & 3.733 & 3.7369 & 3.77  & -- & 0.03986 & 0.00372 & 0     & -- & -6.176 & -11.727 \\
			$^{66} $Ti & 510.176 & 508.051 & 507.01 & -- & 3.778 & 3.7803 & 3.78  & -- & 0.206 & 0.21763 & 0.011 & -- & -3.801 & -2.315 \\
			$^{68} $Ti & 512.949 & 511.006 & 510.5 & -- & 3.798 & 3.7978 & 3.8   & -- & 0.22833 & 0.2319 & 0.128 & -- & -3.474 & -6.918 \\
			$^{70} $Ti & 515.197 & 513.276 & 513.73 & -- & 3.819 & 3.8182 & 3.8   & -- & 0.27777 & 0.26511 & 0     & -- & -2.898 & -7.748 \\
			$^{72} $Ti & 516.597 & 514.828 & 515.85 & -- & 3.831 & 3.8313 & 3.81  & -- & 0.25652 & 0.24713 & 0     & -- & -2.505 & -9.889 \\
			$^{74} $Ti & 517.177 & 515.874 & 515.08 & -- & 3.838 & 3.8411 & 3.82  & -- & 0.18922 & 0.21123 & 0     & -- & -2.234 & -12.311 \\
			$^{76} $Ti & 517.602 & 516.697 & 513.06 & -- & 3.848 & 3.8495 & 3.84  & -- & 0.17842 & 0.16559 & 0     & -- & -1.982 & -13.973 \\
			\hline \hline
		\end{tabular}%
	\end{adjustbox}
	\label{tab2}%
\end{table*}%

\begin{figure}[htpb]
		\includegraphics[scale=0.47]{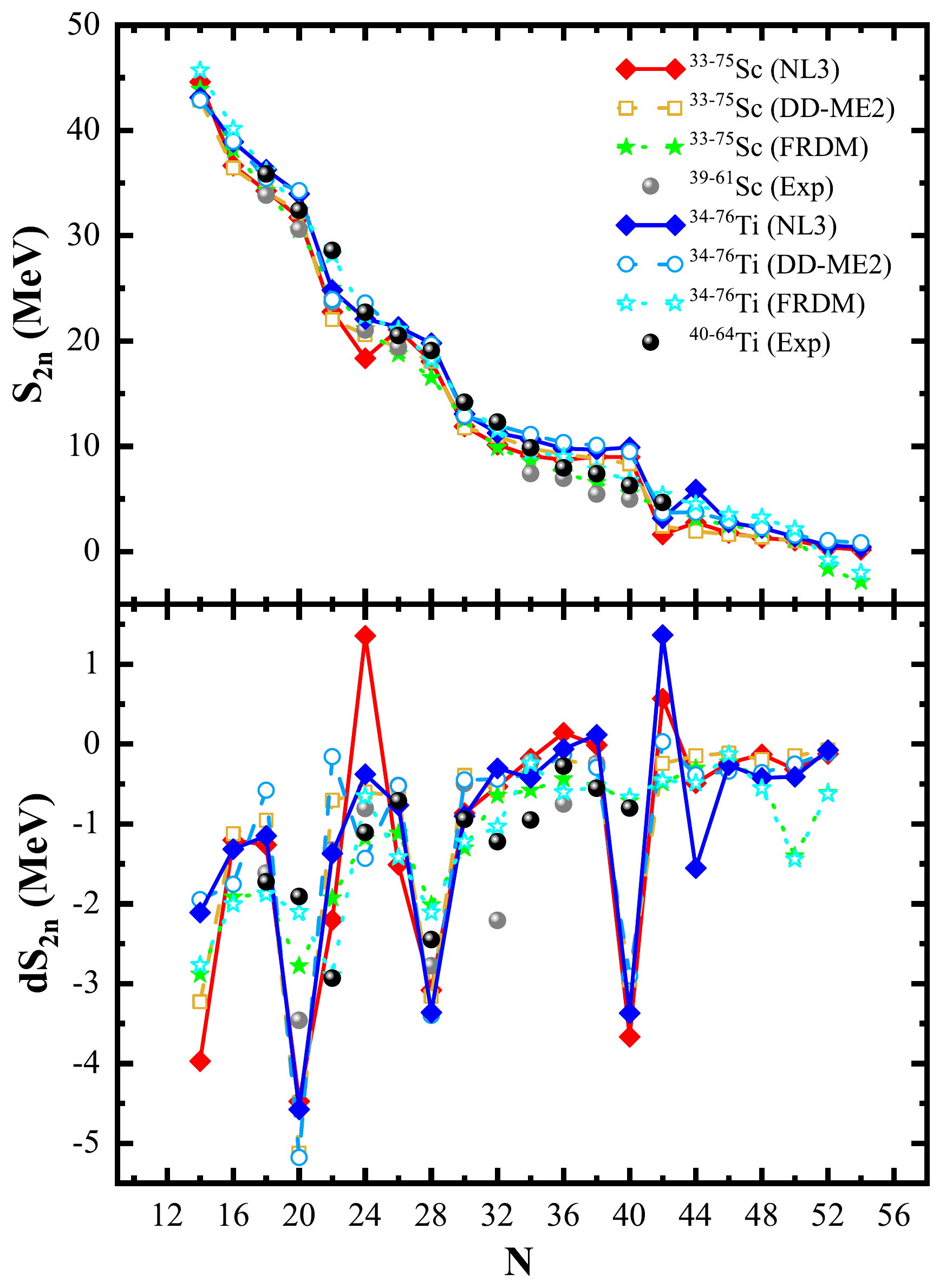}
		\caption{The two neutron separation energies $S_{2n}$ (upper panel) and the differential variation of the separation energy $dS_{2n}$ (lower panel) from the NL3 and DD-ME2 interaction as a function of neutron number ($N$) for Sc and Ti isotopic chains. The FRDM interactions \cite{Moeller2016} and available experimental data \cite{Wang2012} are provided for comparison.}
		\label{fig3}
	\end{figure}
	
\noindent	
{\bf Separation energy and its differential:}
The two-neutron separation energy $S_{2n}$ is defined as the energy needed to remove two neutrons from the nucleus. It is estimated from the ground state nuclear binding energies, $B.E. (Z, N)$ and $B.E. (Z, N-2)$ using the formula, $S_{2n} (Z, N) = - B.E. (Z, N) + B.E. (Z, N-2)$. The two-neutron separation energy provides comprehensive information about the $Q$-value corresponding to a hypothetical simultaneous transfer of two neutrons into the N-2 ground state, resulting from the ground state of the nucleus having $N$ number of neutrons. The two-neutron separation energy of a nucleus explicitly depends on the variations in the ground-state properties of both nuclei, namely configuration, deformation (shape), pairing, and neutron-to-proton asymmetry. Further, the $Q$-value is a critical parameter for determining the possibility of spontaneous and simultaneous emissions of neutrons. 

The $B.E. (N, Z)$ and $B.E. (N-2, Z)$ nuclei are calculated using the non-linear NL3 and density-dependent DD-ME2 interaction parameters. For predicting the correct estimate of nucleon separation energy $S_{2n}$, it is necessary to predict precise mass measurements. In the top panel of Fig. \ref{fig3}, we have compared the $S_{2n}$ as a function of neutron number for \textit{odd-A} isotopes of Scandium ($Z$ = 21) and \textit{even-even}  isotopes of Titanium ($Z$ = 22) nuclei for both parameters. The FRDM predictions \cite{Moeller2016} and latest experimental data \cite{Wang2012} are also given for comparison. In Fig. \ref{fig3}, it can be observed that in an isotopic chain, the $S_{2n} $ decreases with increasing neutron number. A few significant kinks appear at neutron magic numbers $N$ = 20, 28, 40, and an inconsequential observation for $N$ = 50 for both the isotopic chains within NL3 and DD-ME2 parameter sets.  These kinks in $S_{2n}$ values justify the stability of the nucleus, i.e., the energy required for the removal of two neutrons from the nucleus with the neutron number $N_{magic} \pm 2$ is smaller in magnitude than the energy needed for the removal of two neutrons from the nucleus associated with $N_{magic}$. Here, $N_{magic}$ refers to magic number configuration of neutrons.    

The differential variation of two-neutron separation energy ($dS_{2n}$) as a function of neutron number ($N$) is given as, $dS_{2n}(Z,N)=\frac{S_{2n}(Z,N+2)-S_{2n}(Z,N)}{2}$. The $dS_{2n}$ is a parameter used for exploring the rate of change of separation energy with respect to the change in neutron number along an isotopic chain. In the bottom panel of Fig. \ref{fig3}, we have compared the calculated $ dS_{2n} $ for NL3 and DD-ME2 parameters with the FRDM predictions \cite{Moeller2016}  and the latest available experimental data \cite{Wang2012}. Generally, a large, deep, and sharp fall in the $ dS_{2n} $ predicts the existence of shell closure in the isotopic chain, providing much-needed additional information about the nuclear structure. These sharp discontinuities are observed at neutron magic numbers $N$ = 20 and $N$ = 28, with a marginal kink at $N$ = 50. At the same time, large kinks are also observed at $N$ = 24 and 40.  The peak trends observed in $dS_{2n}$ values are similar to that of $ S_{2n} $ values for the respective Scandium and Titanium isotopic chains of nuclei. It is observed that the calculated $ S_{2n} $ and $ dS_{2n} $ are in agreement with the theoretically predicted FRDM \cite{Moeller2016} and experimentally available data \cite{Wang2012}.
	
\begin{figure}[htpb]
	\includegraphics[width=8.4cm,height=11.5cm]{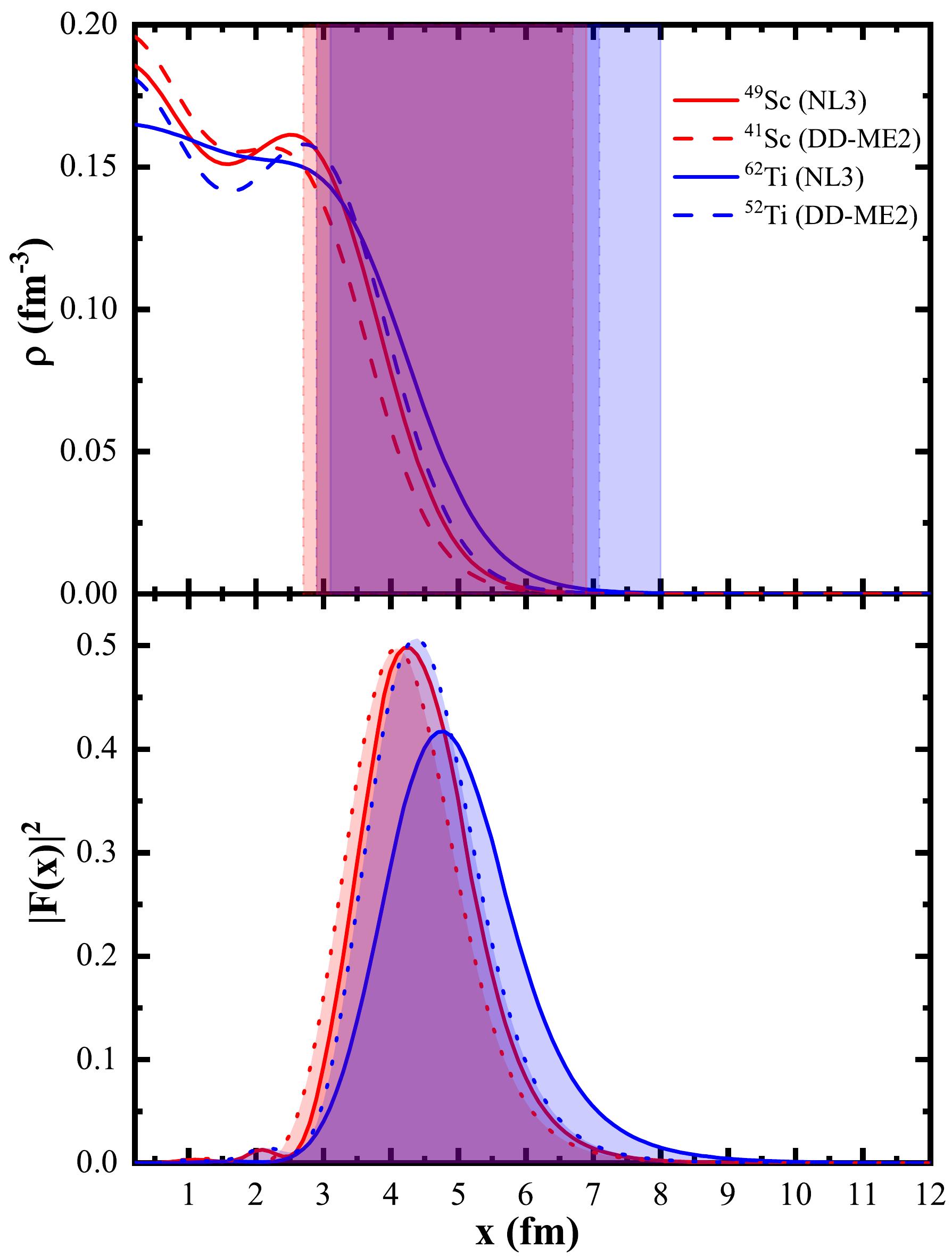}
	\caption{Total density distribution (upper panel) and corresponding weight function (lower panel) for $^{49}$Sc (NL3), $^{62}$Ti (NL3), $^{41}$Sc (DD-ME2) and $^{52}$Ti (DD-ME2) isotopes. Follow the text for details.}
		\label{fig5}
\end{figure}
{\bf \subsection{Isospin properties of finite nuclei}}
\label{isospin}\noindent
{\bf Nuclear density distribution and weight function:}
The isospin properties, namely symmetry energy and its components, are connected with the surface characteristics of nuclear density distributions \cite{antonov1979model,antonov1982spectral,Sarriguren2007,Gaidarov2012,Gaidarov2011}. To obtain the symmetry energy of finite nuclei, we need to follow two steps, firstly, using the density distribution obtained from the RMF formalism with non-linear NL3 and Relativistic-Hartree-Fock with density-dependent DD-ME2 parameter sets, we generate the weight function $\vert \mathcal{F} (x) \vert^{2}$ using Eq. (\ref{eqn:17}) for each nucleus \cite{antonov1979model,antonov1982spectral,Sarriguren2007,Gaidarov2012,Gaidarov2011}. In the second step, we use the weight function and corresponding saturation properties of nuclear matter to calculate the effective symmetry energy in the realm of finite nuclear matter \cite{antonov1979model,antonov1982spectral,Sarriguren2007,Gaidarov2012,Gaidarov2011}.

In Fig. \ref{fig5} (upper panel), we present the total density distribution ($\rho$), i.e.,  the sum of the neutron density ($\rho_{n}$) and the proton density ($\rho_{p}$) as a function of nuclear distance for \textit{odd-A} Sc- and \textit{even-even} Ti- isotopes for NL3 and DD-ME2 parameter sets. A thorough investigation of the graphs shows that with increasing proton number (\textit{Z}), minute enhancement in the surface region is observed. Thus, the total density distribution has a profound role in the effective nuclear matter quantities. The weight function $\vert \mathcal{F} (x) \vert^{2}$, which is calculated from the corresponding total densities of the nuclei, is displayed in the lower panel of Fig. \ref{fig5}. By comparing the upper and lower panel, it can be inferred that the surface density profile is reflected in the weight function. In other words, the value of lower central density provides a lesser height of the weight function for given nuclei and maxima at the surface. \\
	
The $\vert \mathcal{F}(x) \vert^{2}$ in Eq. (\ref{eqn:17}) has a bell-shaped form with a maximum value is observed near $ x=R_{1/2} $, where the density value $ \rho (R_{1/2})$ is about half the value of central density $\rho_{0}$. In other words, the region surrounding $\rho(R_{1/2}) = 0.5 \times \rho_{0}$, is nothing but the surface region of the nuclear matter density distribution. In principle, following Eq. (\ref{eqn:18}), the integration limits are set from 0 to $\infty$. However, using the Brueckner energy density functional method, it is observed that symmetry energy has non-physical values in some regions. Hence, it is crucial to set the proper limits of the integration, i.e., $x_{min}$ and $x_{max}$, where the symmetry energy at local density $ S^{NM}(x) $ changes sign from negative to a positive value and vice-versa respectively. It must be noted that no point of $ S^{NM}(x) $ corresponding to large $ x $ has been observed to change the sign from positive to negative. Instead, $S^{NM} (x)$ for large $x$ tends to zero. Thus, $x_{max}$ is introduced in the right part of the weight function $\vert \mathcal{F} (x) \vert^{2}$, beyond which the contribution to symmetry energy (from $ x_{max} $ to $\infty$) is negligible. The estimated values of $x_{min}$ and $x_{max}$ for the specified isotopes of Sc- and Ti- isotopes for NL3 (solid line) and DD-ME2 (dotted line) parameter sets are filled with light red, and light blue color, respectively. It is worth mentioning that a considerable part of the weight functions attains the peak in the range corresponding to the surface region of the density distributions. This infers the importance of the surface contribution of density while calculating the symmetry energy, and hence these quantities are also known as surface properties. 
	\begin{figure}[htpb]
		\includegraphics[scale=0.30]{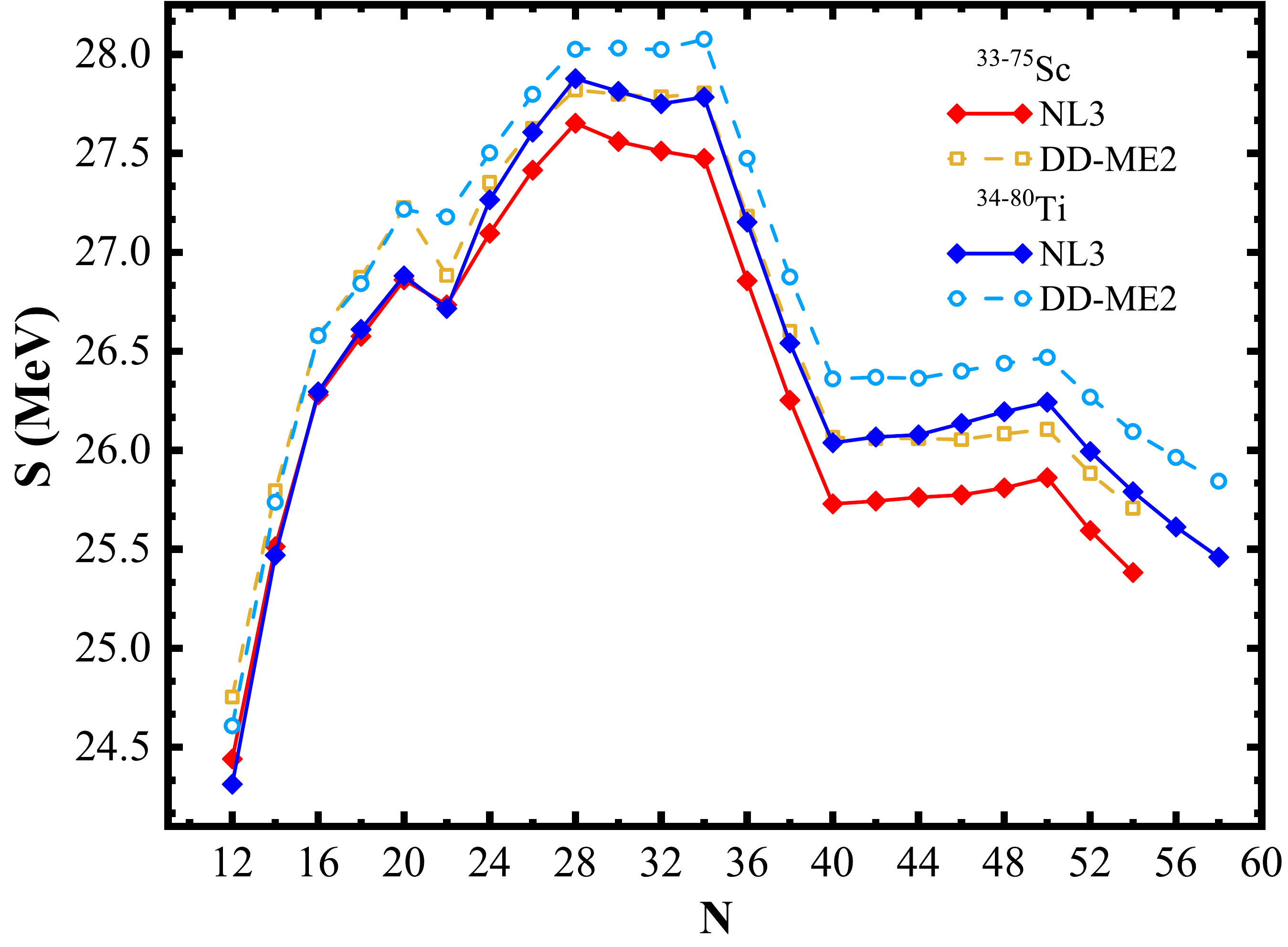}
		\caption{The symmetry energy $S$ as a function of neutron number $N$ for the NL3 and DD-ME2 interactions are given for Sc and Ti isotopic chains. Follow the text for details.}
		\label{fig7}
	\end{figure}
	\begin{figure}[htpb]
		\includegraphics[scale=0.30]{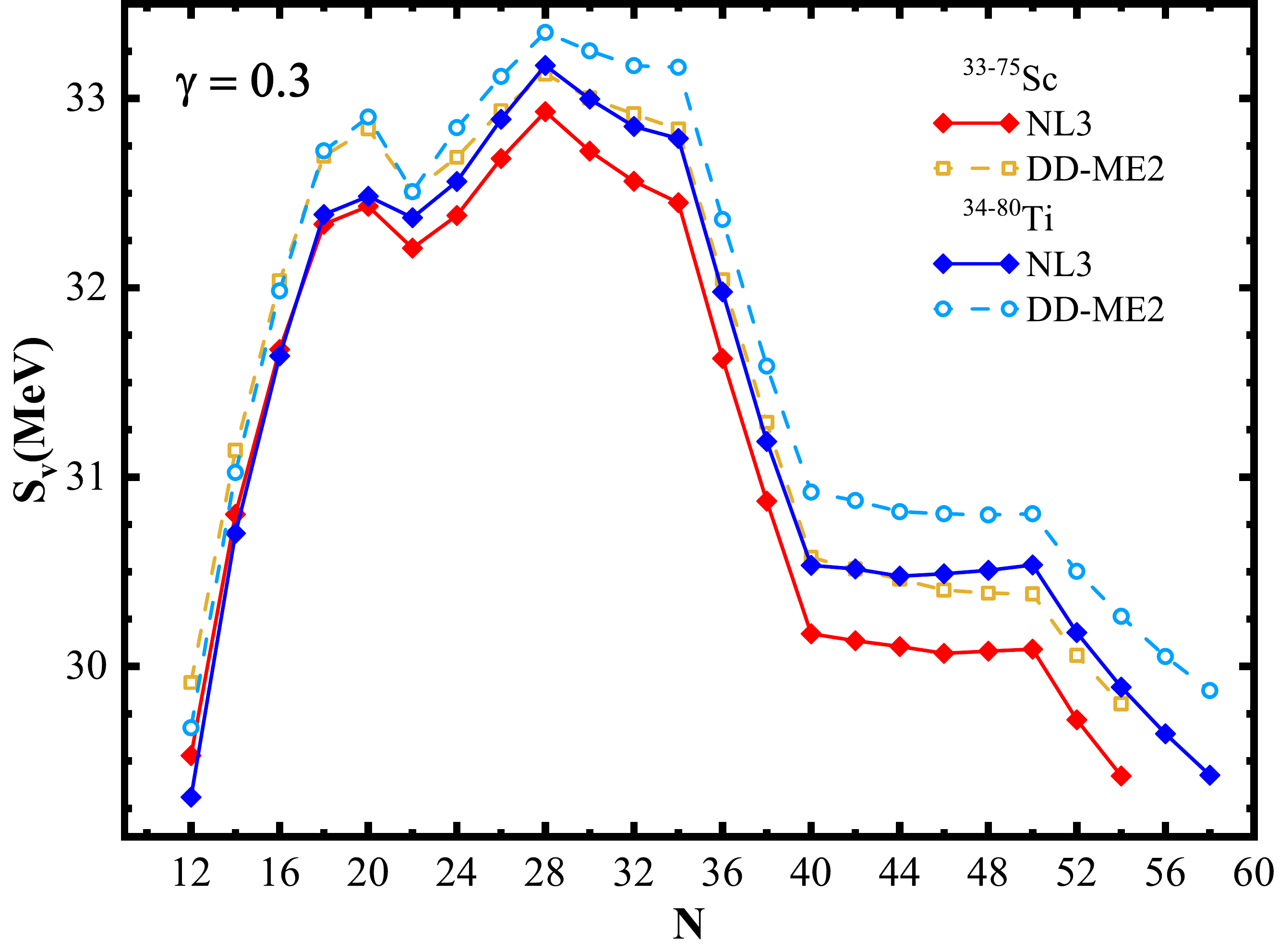}
		\caption{The volume symmetry energy $S_{V}$ as a function of neutron number $N$ for the NL3 and DD-ME2 interactions are given for Sc and Ti isotopic chains. Follow the text for details.}
		\label{fig9}
	\end{figure}
	\begin{figure}[htpb]
		\includegraphics[scale=0.30]{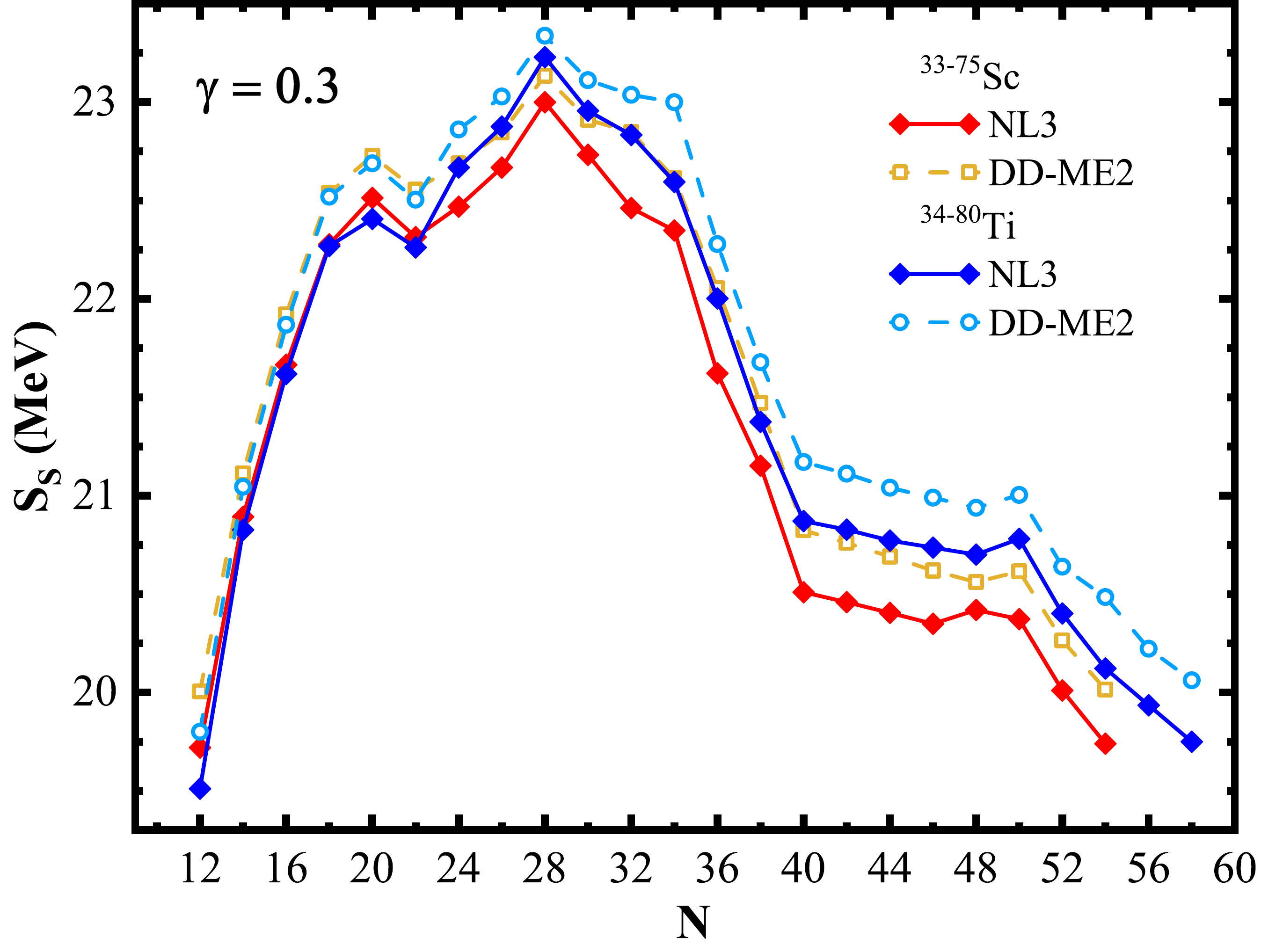}
		\caption{The surface symmetry energy $S_{S}$ as a function of neutron number $N$ for the NL3 and DD-ME2 interactions are given for Sc and Ti isotopic chains. Follow the text for details.}
		\label{fig10}
	\end{figure}
	\begin{figure}[htpb]
		\includegraphics[scale=0.30]{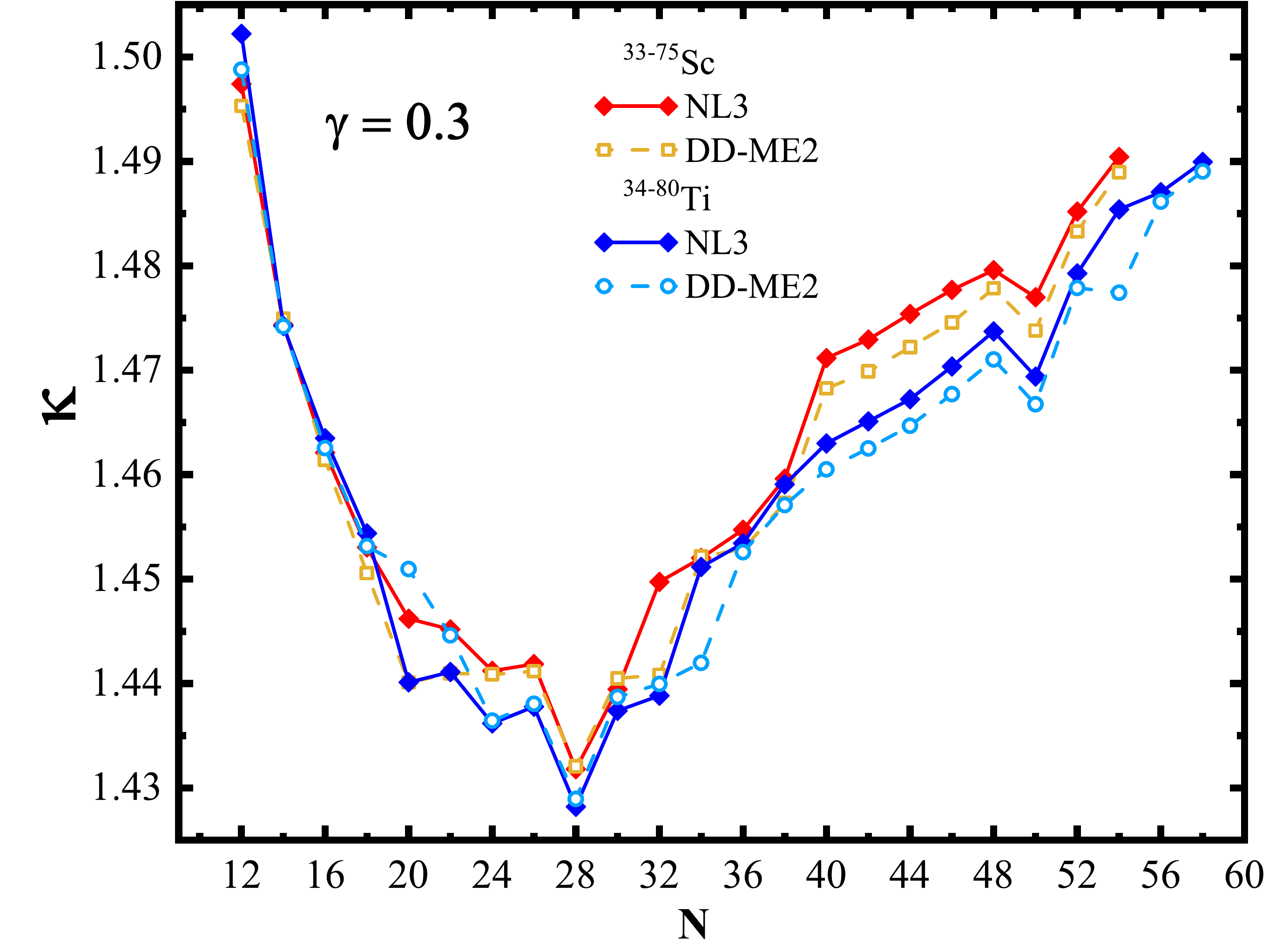}
		\caption{The $\kappa$ for isotopes a function of neutron number $N$ for the NL3 and DD-ME2 interactions are given for Sc and Ti isotopic chains. Follow the text for details.}
		\label{fig11}
	\end{figure}
	\begin{figure}[htpb]
		\includegraphics[width=8.3cm,height=14cm]{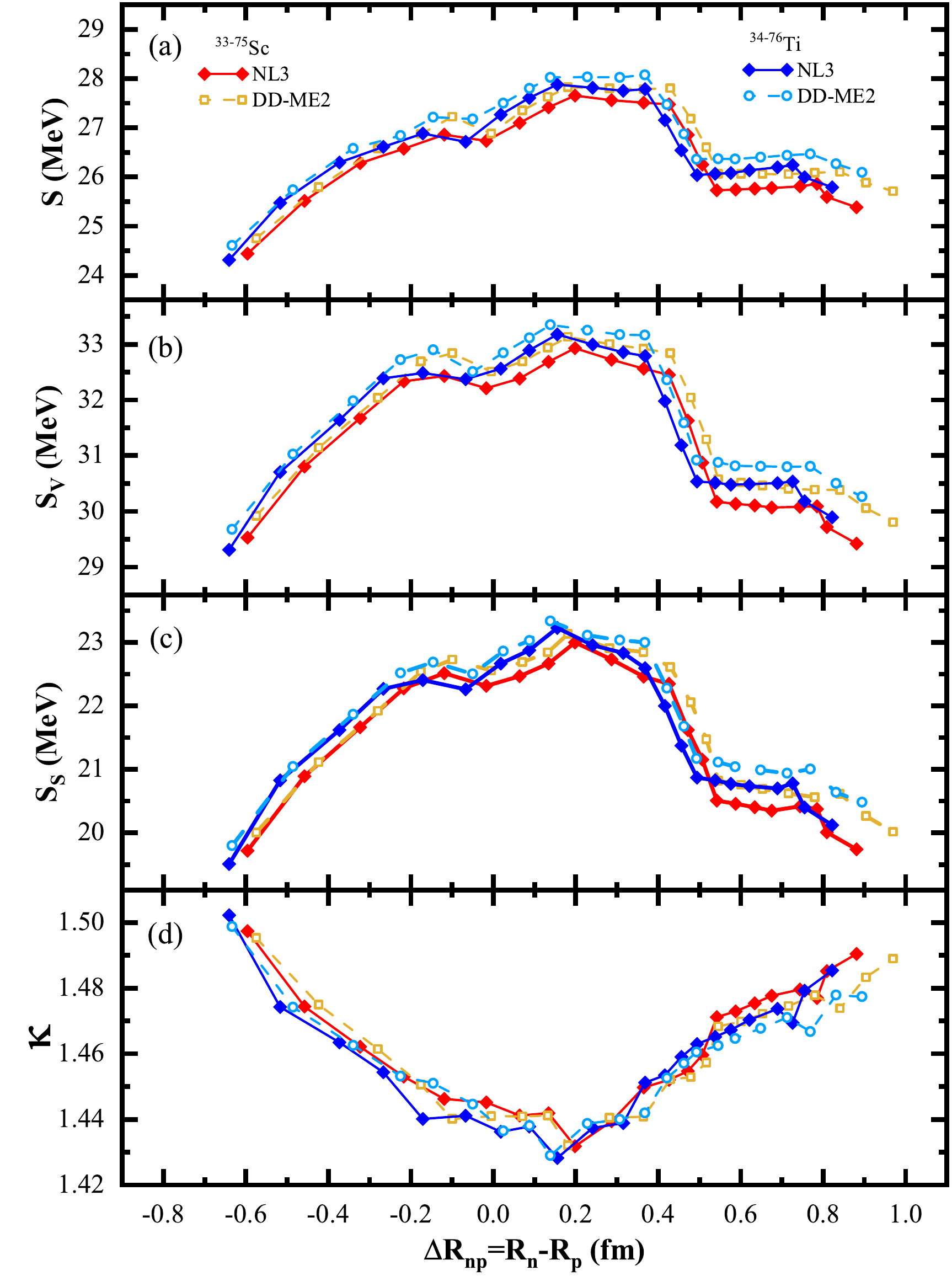}
		\caption{The (a) symmetry energy $S$, (b) volume symmetry energy $S_{V}$, (c) surface symmetry energy $S_{S}$ and (d) their ratio $\kappa$ for isotopic chain of Sc- and Ti- nuclei as a function of neutron-skin thickness $ \Delta R_{np} $ for the NL3 and DD-ME2 parameter sets. Follow the text for details.}
		\label{fig12}
	\end{figure}	
	
\noindent
{\bf Symmetry energy:} As we move across the nuclear landscape from the stability line towards the drip line, the traditional observables fail to provide the relevant signature of shell and/or sub-shell closure. Symmetry energy being a quantity dependent on isospin asymmetry is one of the most critical properties capable of determining the magicity of the nuclei near and beyond the drip-line region (Ref. \cite{Bhuyan2018} and reference therein). Moreover, the symmetry energy coefficient of a nucleus can be described in terms of the coefficients of surface and volume symmetry energy \cite{Antonov1994}. In Ref. \cite{Nikolov2011}, it is observed that the volume-symmetry energy is nearly shape-independent quantity, which infers that the surface effects do not affect the volume-symmetry energy. The surface effects become negligibly small for heavy mass nuclei as its surface symmetry energy coefficient is proportional to $ A^{-1/3} $ \cite{Nikolov2011,Agrawal2012}. Thus, the surface effects are significant for lighter nuclei and have little implication for heavier mass nuclei. Taking deformations into account may provide better results. Recently, the effect of deformation of finite nuclei on the symmetry energy using the theoretical Thomas–Fermi approximation over Skyrme energy density functional has been reported \cite{Mo2015}, which highlights the effect of deformation on symmetry energy at local density decreases with increasing mass number. Here we also find that the relative change in the symmetry energy is about 0.4 MeV for a relatively large value of $\beta_{2} \approx 0.6$. Thus, we have only considered the spherical densities of these isotopic chains of nuclei for computational ease in our calculations.

The symmetry energy is calculated using Eq. (\ref{eqn:18}) for Sc- and Ti- isotopic chain based on non-linear NL3 and density-dependent DD-ME2 parameter sets. The results are shown in Fig. \ref{fig7}. As can be observed from the figure, numerous sharp discontinuities or kinks are observed at neutron magic numbers corresponding to $N$ = 20, 28 and 50, signifying shell/sub-shell closures for each isotopic chain. We did not find any signature of magicity for $N$ = 40 over the isotopic chain of both nuclei. Relatively minor discontinuities are also observed at $N$ = 34. These kinks exhibit that these nuclei have higher stability at the magic neutron number than their surrounding isotopes in the corresponding isotopic chain of nuclei. In other words, abnormalities observed in the symmetry energy curve suggest that additional energy would be needed to convert one neutron to a proton or vice versa. The symmetry energy $S$ observed for the DD-ME2 parameter shows a consistently more significant value than the NL3 parameter for both the isotopic chain. It is to be noted that the nuclear matter symmetry energy for DD-ME2 typically has a higher magnitude than the NL3 parameter set. Hence, it shows that the surface contribution from DD-ME2 dominates the nuclear matter symmetry energy value in finite nuclei. As a result, we get a slightly significant value of symmetry energy for DD-ME2 than NL3 for both the isotopic chain at local density. 
	
\noindent	
{\bf Volume and surface symmetry energy:}
As discussed earlier, the components of symmetry energy of a nucleus can be represented in terms of the coefficients of surface and volume symmetry energy \cite{Antonov1994}. Here we have investigated the volume and surface symmetry energy along with their ratio $\kappa$ with respect to neutron number. We choose the value of $\gamma=0.3$ (see Eq. \ref{eqn:29}), due to the conditions enforced on the analyzed quantities at T=0 MeV is consistent with the available empirical predictions  \cite{danielewicz2003surface,Danielewicz2006,Dieperink2007}. For $\gamma = 0.3$, the calculated values corresponding to; (i) the NL3 parameter set yields the symmetry energy value as 24.4 $\le$ $ S $ $\le$ $ 27.9$ MeV, volume symmetry energy value as 29.5 $\le$ $ S_{V} $ $\le$ $ 33.2 $ MeV and surface symmetry energy value as 19.7 $\le$ $ S_{S} $ $\le$ $ 23.2 $ MeV and (ii) the DD-ME2 parameter set provides the symmetry energy value as 24.3 $\le$ $ S $ $\le$ $ 28.1 $ MeV, volume symmetry energy value as 29.3 $\le$ $ S_{V} $ $\le$ $ 33.4 $ MeV and surface symmetry energy value as 19.5 $\le$ $ S_{S} $ $\le$ $ 23.4 $ MeV. The values obtained using $\gamma = 0.3$ for $S$, $ S_{V}$ and $S_{S}$ are consistent with the predictions of Ref. \cite{danielewicz2003surface}. 
	
The plots of different components of symmetry energy, namely volume symmetry energy, surface symmetry energy, and their ratio ($\kappa$) for Scandium and Titanium isotopes within non-linear NL3 and density-dependent DD-ME2 parameter sets are given in Figs. \ref{fig9}, \ref{fig10} and \ref{fig11} respectively. In all three figures, we can infer that a few sharp discontinuities/kinks are observed at neutron numbers corresponding to traditional magic numbers $N$ = 20, 28 and 50. A piece of limited evidence for neutron magic number at $N$ = 40 is observed for both the isotopic chain within both the parameter sets. Minor kink is also noticed at neutron number $N$ = 34, which is relatively small in magnitude as compared to the relative change at $N$ = 20, 28 and 50. It is worth mentioning that $N$ = 34 is experimentally identified neutron shell/sub-shell closure for this region \cite{steppenbeck2013evidence}. We have also calculated the ratio of volume symmetry energy and surface symmetry energy, denoted as $\kappa$, which follows nearly the same pattern exhibiting considerable depth at $N$ = 20,  28 and 50. These depths in the ratio of volume symmetry energy to the surface component of symmetry energy are consistent with the observed kinks in the symmetry energy. \\
	
\noindent
{\bf Neutron skin thickness and symmetry energy:}
\label{subsect_skin_thick} 
The density-dependence of symmetry energy is the effective source of uncertainty in the equation of state of asymmetric nuclear matter and reasonably constrained only around the saturation density by the bulk properties of the finite nuclei.  At saturation, the density appears to be well correlated with the neutron-skin thickness $\Delta R_{np}$ in the finite nuclear system. Furthermore, recent studies (Ref. \cite{Bhuyan2018} and reference therein) have shown the correlation of neutron-skin thickness with the surface properties of the nuclei and the emergence of kinks at shell/sub-shell closure over an isotopic chain. We show the symmetry energy and its components, namely volume and surface symmetry energy, along with their ratio $\kappa$ as a function of neutron skin thickness ($\Delta R_{np} $) for the Sc- and Ti- isotopic chains based on NL3 and DD-ME2 parameter sets in Fig. \ref{fig12} (a), (b), (c), and (d) respectively. It is noticed that the value of symmetry energy decreases with neutron numbers in either direction of the magic or semi-magic number in the isotopic chain. For example, we find kink (depth) at $N$ = 20, 28, 34, and  50 for the symmetry energy and its components over the isotopic chains of Sc- and Ti- nuclei for both the parameter sets, which further strengthen the confirmation of the above predictions. The appearance of a kink over an isotopic chain shows that the energy required to convert one proton-to-neutron is significantly more for the shell or sub-shell closure isotope than its neighboring nuclei. \\
	
\section{Summary and Conclusions}
\label{Summary}
In the present study, we have investigated the symmetry energy along with its components for \textit{odd-A} Scandium and \textit{even-even} Titanium isotopic chains. The relativistic mean-field with non-linear NL3 and the Relativistic-Hartree-Bogoliubov approach with density-dependent DD-ME2 interaction parameter sets are employed. The coherent density fluctuation model is adopted to correlate the infinite nuclear matter quantities existing in momentum space to their corresponding finite nuclear quantities present in the coordinate space. We have also probed the ground state bulk properties such as the binding energy ($B.E.$), \textit{rms} charge radius ($r_{ch} $), quadrupole deformation ($\beta_{2}$), two-neutron separation energy ($S_{2n} $) and differential variation of two-neutron separation energy ($dS_{2n}$) for the traditional structural analysis. Furthermore, we observe a shape transition from spherical to prolate for $N$ $\geq$  44 and $N$ $\geq$ 40 for the Sc- and Ti- isotopic chain, respectively.

The effective surface and isospin-dependent nuclear matter quantities such as symmetry energy, volume and surface components, are determined for finite nuclei at local densities. We found sharp discontinuities or kinks in the symmetry energy, along with volume and surface symmetry energy corresponding to neutron numbers, $N$ = 20,  28, 34, and 50. It is worth mentioning that the signature/evidence of shell/sub-shell closure for both the isotope corresponding to $N$ = 34 and 50 appears only in the isospin-dependent quantities, namely symmetry energy and its components, contrary to the traditional observables. Again, the present analysis strengthens the previous works on various \textit{even-even} nuclei based on different theoretical models in Refs. \cite{Bhuyan2018,Kaur2020,Antonov2016}. We did find a narrow window of magicity at $N$ = 40 for both the isotopic chain. Furthermore, a minute signature of magicity/shell closure appears at $N$ = 34 for both the isotopic chain, which is consistent with the experimental evidence for this region \cite{steppenbeck2013evidence}. Quantitative comparison among the two-parameter used shows that the magnitude of symmetry energy and its components for DD-ME2 is slightly greater than NL3 parameter sets for both the isotopic chain, which is precisely the opposite of the saturation value of symmetry energy in infinite nuclear matter. Hence, in contrast to the traditional observables, the isospin-dependent surface properties of nuclei in terms of weight function play a crucial role in theoretical confirmation and prediction of newer shell/sub-shell closure over the isotopic chain of finite nuclei at local density. 
	
\section*{Acknowledgements}
	This work has been supported by Science and Engineering Research Board (SERB), Department of Science and Technology (DST), Govt. of India, File No. CRG/2021/001229; FOSTECT Project Code.: FOSTECT.2019B.04; FAPESP Project No. 2017/05660-0, and INCT-FNA Project No. 464898/2014-5, and NKFIH (K134197).

\end{document}